\documentclass[useAMS,usenatbib]{mn2e}

\usepackage{graphicx,amssymb,multirow,color}
\usepackage[fleqn]{amsmath}  

\bibpunct{(}{)}{;}{a}{}{,}
\usepackage{aas_macros} 

\def\simlt{\lower.5ex\hbox{$\; \buildrel < \over \sim \;$}}
\def\simgt{\lower.5ex\hbox{$\; \buildrel > \over \sim \;$}}

\def\eg{{\it e.g.}}

\usepackage{graphicx} 
\newcommand{\be}{\begin{equation}}
\newcommand{\ee}{\end{equation}}
\newcommand{\ba}{\begin{eqnarray}}
\newcommand{\ea}{\end{eqnarray}}

\title[How well can CTI be corrected?]{How well can Charge Transfer Inefficiency be corrected?\\
A parameter sensitivity study for iterative correction}
\author[Israel et al.]{Holger Israel$^{1,2,*}$, Richard Massey$^{1,3}$, Thibaut Prod'homme$^{4}$, Mark Cropper$^{5}$
\newauthor Oliver Cordes$^{6}$, Jason Gow$^{7}$, Ralf Kohley$^{8}$, Ole Marggraf$^{6}$, Sami Niemi$^{5}$,
\newauthor Jason Rhodes$^{9}$, Alex Short$^{4}$, Peter Verhoeve$^{4}$ \\
$^{1}$Institute for Computational Cosmology, Durham University, South Road, Durham DH1 3LE, UK\\
$^{2}$Centre for Extragalactic Astronomy, Durham University, South Road, Durham DH1 3LE, UK\\
$^{3}$Centre for Advanced Instrumentation, Durham University, South Road, Durham DH1 3LE, UK\\
$^{4}$European Space Agency, ESTEC, Keplerlaan~1, 2200AG Noordwijk, The Netherlands\\
$^{5}$Mullard Space Science Laboratory, University College London, Holmbury St Mary, Dorking, Surrey RH5 6NT, UK\\
$^{6}$Argelander-Institut f\"ur Astronomie, Universit\"at Bonn, Auf dem H\"ugel 71, 53121 Bonn, Germany\\ 
$^{7}$e2v Centre for Electronic Imaging, The Open University, Walton Hall, Milton Keynes MK7 6AA, UK\\
$^{8}$European Space Agency, ESAC, P.O. Box 78, 28691 Villanueva de la Ca\~{n}ada, Madrid, Spain\\
$^{9}$Jet Propulsion Laboratory, California Institute of Technology, 4800 Oak Grove Drive, Pasadena, CA 91109, United States\\
$^{*}$e-mail: {\tt holger.israel@durham.ac.uk}} \vspace{-0.5cm}

\begin{document}

\date{Accepted ---. Received ---; in original form \today. \vspace{-3mm}}

\pagerange{\pageref{firstpage}--\pageref{lastpage}} \pubyear{2014}

\maketitle

\label{firstpage}

\begin{abstract}
Radiation damage to space-based Charge-Coupled Device (CCD)
detectors creates defects which result in an
increasing Charge Transfer Inefficiency (CTI) that causes spurious image trailing.
Most of the trailing can be corrected during post-processing, by modelling the 
charge trapping and moving electrons back to where they belong. 
However, such correction is not perfect -- and damage is continuing to accumulate 
in orbit. 
To aid future development, we quantify the limitations of current approaches, 
and determine where imperfect knowledge of model parameters 
most degrade measurements of photometry and morphology.

As a concrete application, we simulate $1.5\times10^{9}$
``worst case'' galaxy 
and $1.5\times10^{8}$ star images
to test the performance of 
the \textit{Euclid} visual instrument detectors. 
There are two separable challenges:
If the model used to correct CTI is perfectly the same as that used to add CTI, 
$99.68$~\% of spurious ellipticity is corrected in our setup. 
This is because readout noise is not subject to CTI, 
but gets over-corrected during correction. Second, if we assume the first
issue to be solved, knowledge of the charge trap density within
$\Delta\rho/\rho\!=\!(0.0272\pm0.0005)$\%,
and the characteristic release time of the dominant species to be known within
$\Delta\tau/\tau\!=\!(0.0400\pm0.0004)$\% will be required. 
This work presents the next level of definition of 
in-orbit CTI calibration procedures for \textit{Euclid}.
\end{abstract}


\section{Introduction}
The harsh radiation environment above the Earth's atmosphere gradually degrades
all electronic equipment, including the sensitive Charge-Coupled Device (CCD)
imaging detectors used in the {\it Hubble Space Telescope} (HST) and {\it Gaia} 
\citep{2008IAUS..248..217L}, and proposed for use by {\it Euclid} 
\citep{2011arXiv1110.3193L}. 
CCD detectors work by collecting photoelectrons which are stored 
within a pixel created by an electrostatic potential well.
After each exposure these electrons are transferred via a process called 
clocking, where alternate electrodes are held high and low to move charge 
through the pixels towards the serial register.
The serial register is then clocked towards the output circuit where 
charge-to-voltage conversion occurs providing an output signal 
dependent on the charge contained within a pixel.
The amount of charge lost with each transfer is described by the Charge
Transfer Inefficiency (CTI). One of the results of radiation-induced defects 
within the silicon lattice is the creation of charge traps at different energy
levels within the silicon band-gap. These traps can temporarily capture 
electrons and release them after a characteristic delay, increasing the CTI. 
Any electrons captured during charge transfer can re-join a charge packet later,
as spurious charge, often observed as a charge tail behind each source. 

Charge trailing can be (partially) removed during image postprocessing.
Since charge transfer is the last process to happen during data acquisition, 
the fastest and most successful attempts to correct CTI take place 
as the second step of data reduction, right after the
analogue-digital converter bias has been subtracted.
\citep[\eg][]{2003astro.ph.10714B}. By modelling the solid-state physics of the 
readout process in {\it HST}'s {\it Advanced Camera for Surveys} (ACS), then 
iteratively reversing the model, \citet{2010MNRAS.401..371M} demonstrated a 
$10$-fold reduction in the level of charge trailing. 
The algorithm was sped up by \citet{2010PASP..122.1035A} and incorporated into 
STScI's {\it HST} default analysis pipeline \citep{2012AAS...21924101S}.
As the radiation damage accumulated, the trailing got bigger and easier to measure.
With an updated and more accurate {\it HST} model, \citet{2010MNRAS.409L.109M} 
achieved a $20$-fold reduction. In an independent programme for {\it Gaia}, 
\citet{2013MNRAS.430.3078S} developed a model using different underlying 
assumptions about the solid-state physics in CCDs. \citet{2014MNRAS.439..887M}
created a meta-algorithm that could reproduce either approach through a choice
of parameters, and optimised these parameters for {\it HST} to correct $98$\% 
of the charge trailing.

The current level of achievable correction is acceptable for most immediate applications.
However, radiation damage is constantly accumulating in {\it HST} and {\it Gaia}; 
and increasing accuracy is required as datasets grow, and statistical uncertainties shrink.
One particularly challenging example of stringent requirements in future 
surveys will be the measurement of faint galaxy shapes by {\it Euclid}.

In this paper, we investigate the effect of imperfect CTI correction, 
on artificial images with known properties. We add charge
trailing to simulated data using a CTI model $\mathbf{M}$, then correct
the data using a CTI model with imperfectly known parameters, 
$\mathbf{M}+\delta\mathbf{M}$. After each stage, we compare the measured
photometry (flux) and morphology (size and shape) of astronomical sources to
their true (or perfectly-corrected) values. We develop a general model to 
predict these errors based on the errors in CTI model parameters.
We focus on the the most important parameters of a `volume-driven' CTI model:
the density $\rho_{i}$ of charge traps, the characteristic time $\tau_{i}$ in 
which they release captured electrons, and the power law index $\beta$ 
describing how an electron cloud fills up the physical pixel volume.

This paper is organised as follows.
In Sect.~\ref{sec:simulations}, we simulate
\textit{Euclid} images and present our analysis methods. 
In Sect.~\ref{sec:estim}, we address the challenge of measuring an average
ellipticity in the presence of strong noise.
We present our CTI model and measure the
CTI effects as a function of trap release timescale $\tau$ in 
Sect.~\ref{sec:modcorr}. Based on laboratory measurements of an irradiated
CCD273 \citep{2012SPIE.8453E..04E}, we adopt a baseline trap model $\mathbf{M}$ 
for the \textit{Euclid} VIS instrument (Sect.~\ref{sec:euclid}). In this context, 
we discuss how well charge trailing can be removed in the presence 
of readout noise. We go on to present our results
for the modified correction model ($\mathbf{M}+\delta\mathbf{M}$) and
derive tolerances in terms of the trap parameters based on \textit{Euclid} 
requirements. We discuss these results in Sect.~\ref{sec:disc} and conclude in 
Sect.~\ref{sec:conclusion}.

\section{Simulations and data analysis} \label{sec:simulations}

\subsection{Simulated galaxy images} \label{sec:imsim}

\begin{figure*}
\includegraphics[width=150mm]{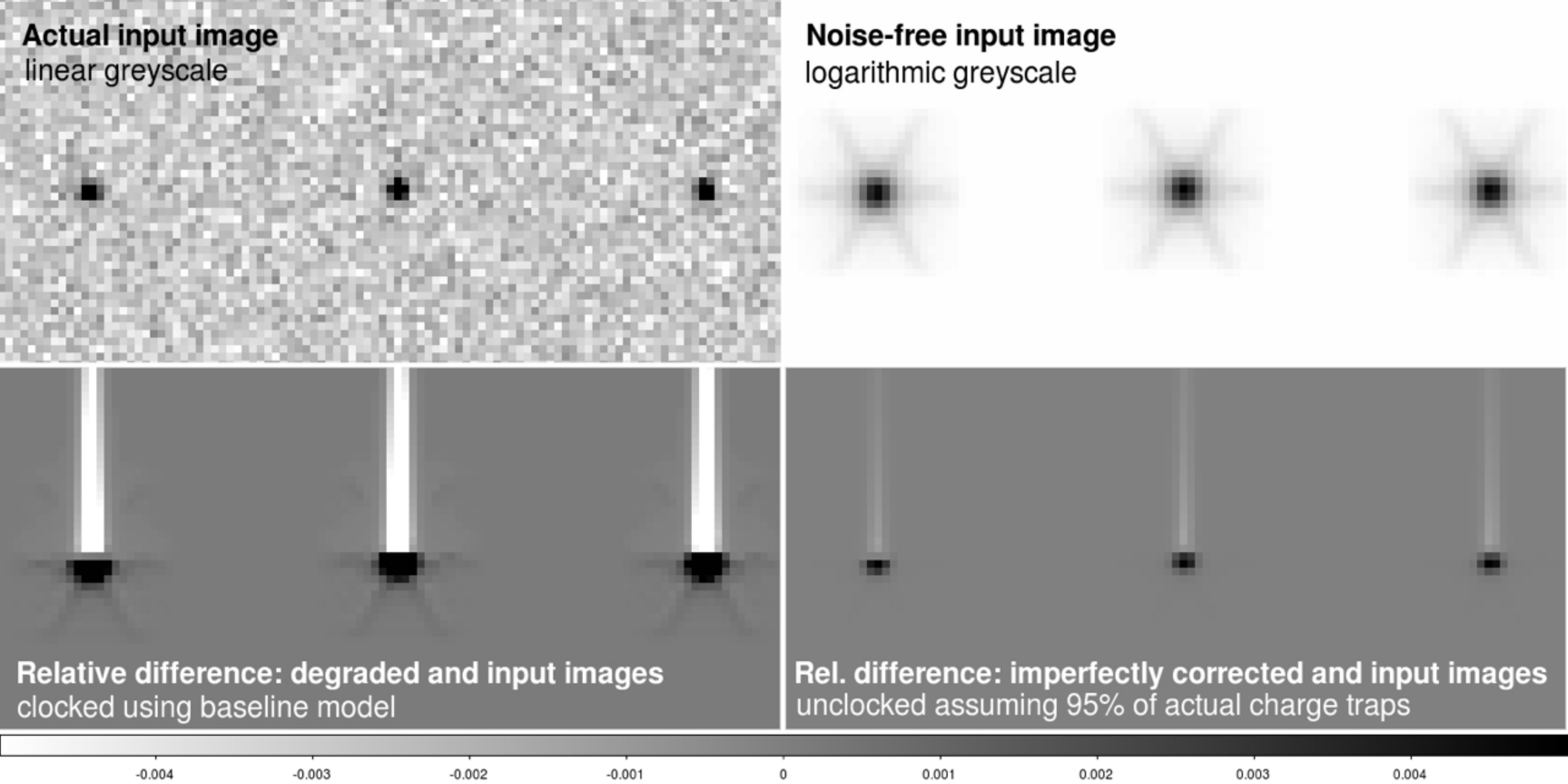}
\caption{Example of three independent noise realisations of our simulated image of a faint galaxy; 
we used one million of these for each measurement. To the input images (\textit{upper left panel}) the clocking is applied,
degrading the image by both CTI and readout noise. The \textit{lower left panel} shows the relative differences
between the degraded and input images for a noise-free realisation (\textit{upper right panel}). We can correct for the CTI,
but not the readout noise (if present), by running the correction software with the trap model as used for clocking.
If the correction model deviates from the clocking model, simulating our
imperfect knowledge of the trap model, the correction will be imperfect. The \textit{lower right panel}
shows the relative difference image for a noise-free example, with the same scale as for the degraded image.
We explore this imperfection; unnoticeable to the naked eye in noisy images, but crucial for \textit{Euclid}'s performance.}
\label{fig:sampleimage}
\end{figure*}

Charge Transfer Inefficiency has the greatest impact on 
small, faint objects that are far from the readout register 
(i.e.\ that have undergone a great number of transfers). 
To quantify the worst case scenario, we therefore simulate the smallest, faintest
galaxy whose properties are likely to be measured -- with an exponential 
flux profile $f(r)\propto\mathrm{e}^{-r}$ whose broad wings (compared to a 
Gaussian or de Vaucouleurs profile) also make it more susceptible to CTI.
To beat down shot noise, we simulate $10^{7}$ noisy image realisations for each measurement.
We locate these galaxies $2048\pm0.5$\,pixels from both the serial readout 
register and the amplifier, uniformly randomising the sub-pixel centre to
average out effects that depend on proximity to a pixel boundary.
All our simulated galaxies have the same circularly symmetric profile, 
following the observation by \citet{2010PASP..122..439R} that this produces the
same mean result as randomly oriented elliptical galaxies with no preferred direction. 

We create the simulated images spatially oversampled by a factor 20, convolve 
them with a similarly oversampled Point Spread Function (PSF), then resample 
them to the final pixel scale. 
We use a preliminary PSF model and the $0\farcs1$\,pixels 
of the \textit{Euclid} VIS instrument, but our setup can easily be
adapted to other instruments, e.g.\ ACS. 
To the image signal of $\sim\!\!1300$ electrons, 
we add a uniform sky background of $105$ electrons,
as expected for a $560\,$s VIS exposure, 
and Poisson photon noise to both the source and the background. 
After clocking and charge trailing (if it is being done; see Sect.~\ref{sec:trailing}),
we then add additional readout noise, which follows a Gaussian distribution with a 
root mean square (rms) of $4.5$\,electrons, the nominal \textit{Euclid} VIS value.

In the absence of charge trailing, the final galaxies have mean $S/N$=$11.35$,
and Full Width at Half Maximum (FWHM) size of $0\farcs18$,
as measured by \texttt{SExtractor} \citep{1996A&AS..117..393B}.
This size, the same as the PSF, at the small end of the range expected from 
Fig.~4 of \citet{2013MNRAS.429..661M} makes our galaxies the most challenging
in terms of CTI correction.
Examples of input, degraded, and corrected images are shown in 
Fig.~\ref{fig:sampleimage}.

Separately, we perform a second suite of simulations,
containing $10^{6}$ realisations of a \textit{Euclid} VIS PSF at
$S/N\!\approx\!200$. The PSF simulations follow the above recipe, 
but skip the convolution of the PSF with an exponential disk.

\subsection{Image analysis}  \label{sec:dataflow}

On each of the sets of images (input, degraded, and corrected), we detect the
sources using \texttt{SExtractor}. Moments of the brightness distribution and
fluxes of the detected objects are measured using an \texttt{IDL}
implementation of the RRG \citep{2001ApJ...552L..85R} shape measurement method.
RRG is more robust than 
\texttt{SExtractor} for faint images, combining Gaussian-weighted moments of 
the image $I(\btheta)$ to measure integrated source flux 
\begin{equation} \label{eq:fmom}
F\equiv\!\int{W(\btheta)\,I(\btheta)\,\mathrm{d}^{2}\btheta},
\end{equation}
where $W(\btheta)$ is a Gaussian weight function with standard
deviation $w$, and the integral extends over $2.5w$;
the position 
\begin{equation} \label{eq:xymom}
y\equiv\!\int{ \theta_2\, W(\btheta)\,I(\btheta)\,\mathrm{d}^{2}\btheta};
\end{equation}
the size 
\begin{equation} \label{eq:rmom}
R^{2}\equiv Q_{11}+Q_{22};
\end{equation}
and the ellipticity
\begin{equation} \label{eq:chidef}
\{e_1,e_2\}\equiv \left\{\frac{Q_{11}-Q_{22}}{Q_{11}+Q_{22}},\frac{2Q_{12}}{Q_{11}+Q_{22}}\right\},
\end{equation}
where the second-order brightness moments are 
\begin{equation} \label{eq:qmom}
Q_{\alpha\beta}\!=\!\int{\theta_{\alpha}\,\theta_{\beta}\,W(\btheta)\,I(\btheta)\,\mathrm{d}^{2}\btheta},
\qquad\{\alpha,\beta\}\!\in\!\{1,2\}.
\end{equation}
For measurements on stars, we chose a window size 
$w\!=\!0\farcs75$, the \textit{Euclid} prescription for stars.
For galaxies, we seek to reproduce the window functions used in weak lensing 
surveys. We adopt the radius of the \texttt{SExtractor} object
\citep[e.g.][]{2007ApJS..172..219L} that with $w\!=\!0\farcs34$
truncates more of the noise and thus returns more robust measurements. 

Note that we are measuring a raw galaxy ellipticity, 
a proxy for the (reduced) shear, in which we are actually interested 
\citep[cf.][for a recent overview of the effects a cosmic shear measurement
pipeline needs to address]{2012MNRAS.423.3163K}. 
A full shear measurement pipeline must also correct ellipticity for convolution by the 
telescope's PSF and calibrate it via a shear `responsivity factor' \citep{1995ApJ...449..460K}. 
The first operation typically enlarges $e$ by a factor of $\sim1.6$ and the 
second lowers it by about the same amount. Since this is already within the 
precision of other concerns, we shall ignore both conversions
The absolute calibration of shear measurement with
RRG may not be sufficiently accurate to be used on future surveys. 
However, it certainly has sufficient {\em relative} accuracy to measure small 
deviations in galaxy ellipticity when an image is perturbed.

\section{High precision ellipticity measurements} \label{sec:estim}

\subsection{Measurements of a non-linear quantity}

A fundamental difficulty arises in our attempt to measure galaxy shapes
to a very high precision, by averaging over a large number of images.
Mathematically, the problem is that calculating ellipticity $e_{1}$ directly from the moments
and then taking the expectation value $\mathcal{E}(\cdot)$ of all objects, i.e.:
\begin{equation} \label{eq:simpleell}
e_{1}=\mathcal{E}\!\left(\frac{Q_{11}-Q_{22}}{Q_{11}+Q_{22}}\right),\quad
e_{2}=\mathcal{E}\!\left(\frac{2Q_{12}}{Q_{11}+Q_{22}}\right),
\end{equation}
means dividing one noisy quantity by another noisy quantity. 
Furthermore, the numerator and denominator are highly correlated.
If the noise in each follows a Gaussian distribution, and their expectation values are zero, the probability
density function of the ratio is a Lorentzian (also known as Cauchy) 
distribution. If the expectation values of the Gaussians are nonzero, as we
expect, the ratio distribution becomes a
generalised Lorentzian, called the Marsaglia-Tin distribution
\citep{Marsaglia65,Marsaglia:2006:JSSOBK:v16i04,Tin65}.
In either case, the ratio distribution has infinite second and first moments, 
i.e.\ its variance -- and even its expectation value -- are undefined. 
Implications of this for shear measurement are discussed in detail by 
\citet{2012MNRAS.424.2757M,2012MNRAS.425.1951R,2012MNRAS.427.2711K,2013MNRAS.429.2858M,2014MNRAS.439.1909V}.

Therefore, we cannot simply average over ellipticity measurements for 
$10^{7}$ simulated images. The mean estimator (Eq.~\ref{eq:simpleell}) 
would not converge, but follow a random walk in which entries from the broad 
wings of the distribution pull the average up or down by an arbitrarily large amount.

\subsection{``Delta method'' (Taylor expansion) estimators for ellipticity}

As an alternative estimator, we employ what is
called in statistics the `delta method': a Taylor expansion of 
Eq.~(\ref{eq:simpleell}) around the expectation value of the denominator
\citep[e.g.][]{casella+berger:2002}. 
The expectation value of the ratio of two random variables $X$, $Y$ 
is thus approximated by:
\begin{multline} \label{eq:deltamethod}
\mathcal{E}(X/Y)\!\approx\!\frac{\mathcal{E}(X)}{\mathcal{E}(Y)}
-\frac{\mathcal{C}(X,Y)}{\mathcal{E}^{2}(Y)}
+\frac{\mathcal{E}(X)\sigma^{2}(Y)}{\mathcal{E}^{3}(Y)}\\
+\frac{\mathcal{C}(X,Y^{2})}{\mathcal{E}^{3}(Y)}
-\frac{\mathcal{E}(X)\mathcal{E}[Y-\mathcal{E}(Y)]^{3}}{\mathcal{E}^{4}(Y)}
\end{multline}
where $\mathcal{E}(X)$, $\sigma(X)$, $\sigma^{2}(X)$ denote the expectation
value, standard deviation, and variance of $X$, and
$\mathcal{C}(X,Y)$ its covariance with $Y$.
The zero-order term in Eq.~(\ref{eq:deltamethod}) is the often-used approximation
$\mathcal{E}(X/Y)\approx\mathcal{E}(X)/\mathcal{E}(Y)$ 
that switches the ratio of the averages for the average of the ratio.
We note that beginning from the first order there are two terms per order
with opposite signs. 
Inserting Eq.~(\ref{eq:qmom}) into Eq.~(\ref{eq:deltamethod}), the first-order
estimator for the ellipticity reads
in terms of the brightness distribution moments $Q_{\alpha\beta}$ as follows:
\begin{align} \label{eq:estim}
\begin{split}
e_{1} =& \frac{\mathcal{E}(Q_{11}\!-\!Q_{22})}{\mathcal{E}(Q_{11}\!+\!Q_{22})} \\
&- \frac{\sigma^{2}(Q_{11})\!-\!\sigma^{2}(Q_{22})}{\mathcal{E}^{2}(Q_{11}\!+\!Q_{22})} +
\frac{\mathcal{E}(Q_{11}\!-\!Q_{22})\sigma^{2}(Q_{11}\!+\!Q_{22})}{\mathcal{E}^{3}(Q_{11}\!+\!Q_{22})}
\end{split}\\
\begin{split}\label{eq:estim_e2}
e_{2} =& \frac{\mathcal{E}(2Q_{12})}{\mathcal{E}(Q_{11}\!+\!Q_{22})}\\
&- \frac{\mathcal{C}(Q_{11},Q_{12})\!+\!\mathcal{C}(Q_{12},Q_{22})}{\mathcal{E}^{2}(Q_{11}\!+\!Q_{22})} +
\frac{\mathcal{E}(2Q_{12})\sigma^{2}(Q_{11}\!+\!Q_{22})}{\mathcal{E}^{3}(Q_{11}\!+\!Q_{22})},
\end{split}
\end{align}

with the corresponding uncertainties,
likewise derived using the delta method
\citep[e.g.][]{casella+berger:2002}:

\begin{align} \label{eq:formal}
\begin{split}
\sigma^{2}(e_{1}) =& \frac{\sigma^{2}(Q_{11}\!-\!Q_{22})}{\mathcal{E}^{2}(Q_{11}\!+\!Q_{22})}\\
&- \frac{\mathcal{E}(Q_{11}\!-\!Q_{22})\left[\sigma^{2}(Q_{11})\!-\!\sigma^{2}(Q_{22})\right]}
{\mathcal{E}^{3}(Q_{11}\!+\!Q_{22})} \\
&+\frac{\mathcal{E}^{2}(Q_{11}\!-\!Q_{22})\sigma^{2}(Q_{11}\!+\!Q_{22})}{\mathcal{E}^{4}(Q_{11}\!+\!Q_{22})}
\end{split} \\
\begin{split} \label{eq:formal_e2}
\sigma^{2}(e_{2}) =& \frac{\sigma^{2}(Q_{11}\!+\!Q_{22})}{\mathcal{E}^{2}(Q_{11}\!+\!Q_{22})}\\
&- \frac{\mathcal{E}(Q_{11}\!+\!Q_{22})\left[\mathcal{C}(Q_{11},Q_{12})\!+
\!\mathcal{C}(Q_{12},Q_{22})\right]}{\mathcal{E}^{3}(Q_{11}\!+\!Q_{22})} \\
&+\frac{\mathcal{E}^{2}(Q_{11}\!+\!Q_{22})\sigma^{2}(Q_{11}\!+\!Q_{22})}{\mathcal{E}^{4}(Q_{11}\!+\!Q_{22})}\quad.
\end{split}
\end{align}

\subsection{Application to our simulations} \label{sec:sigmas}

For our input galaxies, the combined effect of the first-order terms 
in eq.~(\ref{eq:estim}) is $\sim\!10$~\%.
Second-order contributions to the estimator are small, so we truncate
after the first order. However, because of the divergent moments of the
Marsaglia-Tin distribution, the third and higher-order contributions to the
Taylor series increase again.

Nevertheless, while this delta-method estimator neither mitigates noise bias
nor overcomes the infinite moments of the Marsaglia-Tin distribution at a
fundamental level, it sufficiently 
suppresses the random walk behaviour for the purposes of this study, the
averaging over noise realisations of the same object.
We advocate re-casting the \textit{Euclid} requirements
in terms of the \textit{Stokes parameters} 
\citep[$Q_{11}\pm Q_{22},2Q_{12}$;][]{2014MNRAS.439.1909V}. 
These are the numerators and denominator of eq.~(\ref{eq:simpleell})
and are well-behaved Gaussians with finite first and second moments. 

The formal uncertainties on ellipticity we quote in the rest of this article are
the standard errors $\sigma(e_{1})/\sqrt{N}$ given by eq.~(\ref{eq:formal}). 
Our experimental setup of re-using the same simulated sources
(computationally expensive due to the large numbers needed),
our measurements will be intrinsically correlated (Sect.~\ref{sec:corr}).  
Hence the error bars we show overestimate the true uncertainties.

\section{The effects of fast and slow traps} \label{sec:modcorr}

\subsection{How CTI is simulated} \label{sec:trailing}

The input images are degraded using a \texttt{C} implementation of the 
\citet{2014MNRAS.439..887M} CTI model.
During each pixel-to-pixel transfer, in a cloud of $n_{\mathrm{e}}$ electrons, 
the number captured is
\begin{equation} \label{eq:nenc}
n_{\mathrm{c}}(n_{\mathrm{e}}) =
\left(1-\exp{\left(-\alpha n_{\mathrm{e}}^{1\!-\!\beta}\right)}\right)
\sum_{i}\rho_{i}
\left(\frac{n_{\mathrm{e}}}{w}\right)^{\beta},
\end{equation}
where the sum is over different charge trap species with density $\rho_i$ 
per pixel, and $w$ is the full-well capacity. Parameter $\alpha$ controls the 
speed at which electrons are captured by traps within the physical volume of
the charge cloud, which grows in a way determined by parameter $\beta$ .

Release of electrons from charge traps is modelled by a simple exponential
decay, with a fraction $1-\mathrm{e}^{(-1/\tau_{i})}$ escaping during each 
subsequent transfer. The characteristic release timescale $\tau_{i}$ depends on
the physical nature of the trap species and the operating temperature of the CCD.

In this paper, we make the simplifying 
`volume-driven' assumption that charge capture is instantaneous, so $\alpha\!\!=\!\!0$. 
Based on laboratory studies of an irradiated VIS CCD (detailed in 
Sect.~\ref{sec:labdata}), we adopt a $\beta\!=\!0.58$ baseline well fill, 
and end-of-life total density of one trap per pixel, $\rho\!=\!1$.
In our first, general tests, we investigate a single trap species 
and explore the consequences of different values of $\tau$.

\subsection{Iterative CTI correction} \label{sec:corr}
\begin{figure*}
\includegraphics[width=155mm]{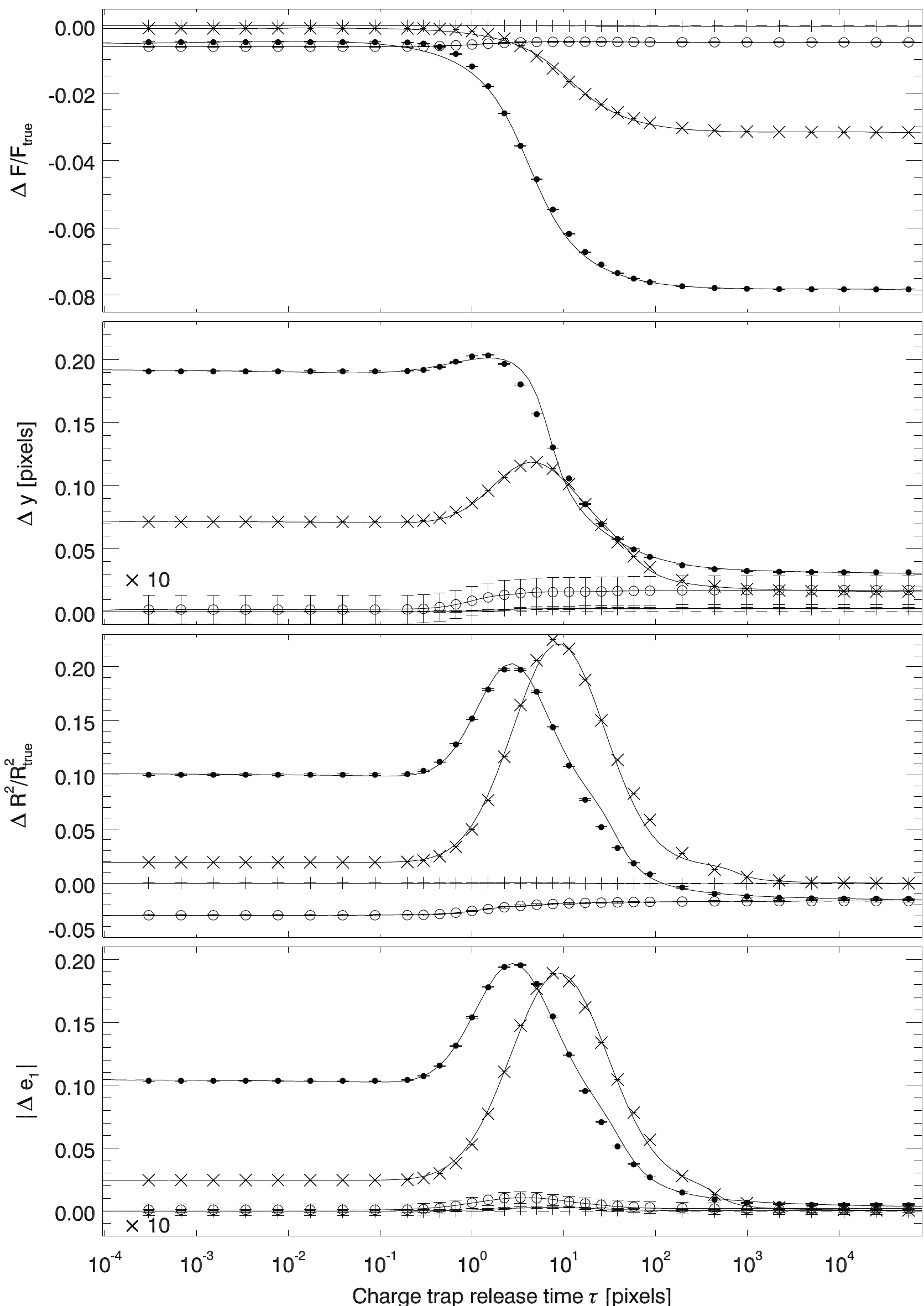}
\caption{The effect on measurements of galaxy flux 
$F$, astrometry $y$ and morphology (size $R^{2}$, ellipticity $e_{1}$) 
of charge traps of different species, i.e.\ release timescale $\tau$.
The pixel-to-pixel transfer is assumed to be instantaneous. 
Filled (open) circles denote data for faint galaxies before (after) CTI mitigation.
Crosses (plus signs) denote data for bright stars before (after) CTI mitigation.
Data for CTI-corrected images are shown 
multiplied by a factor of $10$ for $\Delta y$ and $\Delta e_{1}$. 
Lines give the best fit following Eq.~(\ref{eq:adg}), 
with the coefficients listed in Table~\ref{tab:taufits}.}
\label{fig:raw4panel}
\end{figure*}
\begin{table*}
 \centering
 \caption{Parameters of fitting functions to illustrate the effect on measurements of galaxy 
fluxes $F$ and $F_{\mathrm{S}}$, 
astrometry $y$ and morphology $R^{2}$, $e_1$ of charge traps of different species. 
In all cases, the measurements assume a density of one trap per pixel, and the astrophysical
measurement is fitted as a function of the charge trap's characteristic release time $\tau$ as 
$A+D_a\,{\mathrm{atan}}{((\log{\tau}-D_p)/D_w)}+G_a\exp{((\log{\tau}-G_p)^2/2G_w^2)}$.
Values after correction highlight the efficacy of CTI mitigation.}
\vspace{-1mm}
 \begin{tabular}{lccccccc} \hline\hline
~ & $A$ & $D_{\mathrm{a}}$ & $D_{\mathrm{p}}$ & $D_{\mathrm{w}}$ & $G_{\mathrm{a}}$ & $G_{\mathrm{p}}$ & $G_{\mathrm{w}}$ \\ \hline
\multicolumn{8}{l}{Galaxy simulation: in degraded images, including readout noise $\qquad[10\times$parameter]}\\
$\Delta F/F_{\mathrm{true}}$ & $-0.5367\pm0.0098$ & $-0.3144\pm0.0085$& $6.199\pm0.044$ & $4.639\pm0.260$ & $0.2116\pm0.0194$ & $49.53\pm1.64$ & $41.54\pm2.39$ \\
$\Delta y$ & $1.1098\pm0.0014$ & $-0.5291\pm0.0028$ & $8.392\pm0.080$ & $2.110\pm0.234$ & $0.3061\pm0.0185$ & $6.935\pm0.402$ & $7.083\pm0.210$ \\
$\Delta R^{2}/R^{2}_{\mathrm{true}}$ & $0.4226\pm0.0025$ & $-0.3857\pm0.0038$ & $15.72\pm0.18$ & $2.576\pm0.375$ & $1.0866\pm0.0448$ & $4.382\pm0.047$ & $3.779\pm0.160$ \\
$\Delta e_{1}$ & $0.5333\pm0.0016$ & $-0.3357\pm0.0026$ & $16.28\pm0.22$ & $2.951\pm0.326$ & $0.9901\pm0.0203$ & $4.553\pm0.054$ & $4.132\pm0.081$ \\ \hline
 \multicolumn{8}{l}{Galaxy simulation: after correction in software post-processing (perfect knowledge of charge traps) $\qquad[100\times$parameter]}\\
$\Delta F/F_{\mathrm{true}}$ & $-0.5549\pm0.0029$ & $0.0446\pm0.0028$ & $129.6\pm13.7$ & $26.00\pm13.36$ & $0.1301\pm0.0121$ & $73.47\pm6.78$ & $56.84\pm5.21$ \\
$\Delta y$ & $0.09582\pm0.01011$ & $0.0517\pm0.0111$ & $5.622\pm8.911$ & $2.227\pm4.557$ & $0.0810\pm0.1170$ & $2.757\pm5.369$ & $3.154\pm2.784$ \\
$\Delta R^{2}/R^{2}_{\mathrm{true}}$ & $-2.3181\pm0.0173$ & $0.4431\pm0.0202$ & $75.90\pm25.02$ & $28.47\pm11.03$ & $0.5471\pm0.2294$ & $41.31\pm16.09$ & $35.33\pm9.12$ \\
$\Delta e_{1}$ & $0.01383\pm0.0115$ & $0.0039\pm0.0066$ & $12.30\pm20.49$ & $1.000\pm0.000$ & $0.0982\pm0.0274$ & $5.738\pm2.085$ & $5.353\pm2.078$ \\ \hline
\multicolumn{8}{l}{Star simulation: in degraded images, including readout noise $\qquad[100\times$parameter]}\\
$\Delta F/F_{\mathrm{true}}$ & $-2.2472\pm0.0239$ & $-1.4558\pm0.0189$ & $107.5\pm0.3$ & $55.11\pm0.95$ & $1.151\pm0.047$ & $496.6\pm3.2$ & $343.6\pm4.4$ \\
$\Delta y$ & $4.3532\pm0.0014$ & $-1.8608\pm0.0027$ & $173.1\pm0.4$ & $29.20\pm0.67$ & $5.0987\pm0.0173$ & $67.20\pm0.20$ & $43.91\pm0.22$ \\
$\Delta R^{2}/R^{2}_{\mathrm{true}}$ & $0.9489\pm0.00098$ & $-6.434\pm0.0095$ & $288.8\pm4.7$ & $18.71\pm4.49$ & $20.237\pm0.716$ & $94.42\pm0.15$ & $50.20\pm0.25$ \\
$\Delta e_{1}$ & $1.2336\pm0.0077$ & $-0.7941\pm0.0086$ & $266.7\pm2.4$ & $17.54\pm3.90$ & $16.513\pm0.046$ & $94.87\pm0.19$ & $52.57\pm0.21$ \\ \hline
 \multicolumn{8}{l}{Star simulation: after correction in software post-processing (perfect knowledge of charge traps) $\qquad[100\times$parameter]}\\
$\Delta F/F_{\mathrm{true}}$ & $-0.0035\pm0.0002$ & $0.0027\pm0.0003$ & $110.2\pm10.5$ & $42.21\pm20.02$ & $0.0006\pm0.0271$ & $182.6\pm71.3$ & $3.5\pm100.0$ \\
$\Delta y$ & $0.1504\pm0.00066$ & $0.0970\pm0.0067$ & $12.46\pm1.86$ & $2.731\pm1.552$ & $0.0218\pm0.0034$ & $7.377\pm1.024$ & $5.063\pm0.717$ \\
$\Delta R^{2}/R^{2}_{\mathrm{true}}$ & $-0.0163\pm0.0038$ & $-0.0182\pm0.0036$ & $1269\pm33$ & $24.57\pm47.63$ & $0.0198\pm0.0146$ & $50.83\pm34.56$ & $37.95\pm38.64$ \\
$\Delta e_{1}$ & $0.0012\pm0.0024$ & $0.0003\pm0.0014$ & $2.26\pm50.92$ & $1.000\pm0.000$ & $0.02668\pm0.0061$ & $8.465\pm1.800$ & $5.379\pm1.647$ \\
  \hline\label{tab:taufits}
  \vspace{-4mm}
 \end{tabular}
\end{table*}
The \citet{2014MNRAS.439..887M} code can also be used to `untrail' the CTI.
If required, we use $n_\mathrm{iter}=5$ iterations to attempt to correct the image 
(possibly with slightly different model parameters).
Note that we perform this correction only after adding readout noise 
in the simulated images. 

Our main interest in this study is the impact of uncertainties in the trap
model on the recovered estimate of an observable $\eta$ (e.g.\ ellipticity). 
Therefore, we present our results in terms of 
differences between the estimators measured for the corrected images, and the input values:
\begin{equation} \label{eq:deltadef}
\Delta\eta_{i} = \eta_{i,\mathrm{corrected}} - \eta_{i,\mathrm{input}}.
\end{equation}
Because for each object of index $i$ the noise in the measurements of
$\eta_{i,\mathrm{corrected}}$ and $\eta_{i,\mathrm{input}}$ are strongly correlated, they
partially cancel out. Thus the actual uncertainty of each $\Delta\eta_{i}$ is lower 
than quoted. Moreover, because we re-use the same noise realisation in all our
measurements (cases of different $\rho_{i}$ and $\tau_{i}$), these measurements are
correlated as well.

\subsection{CTI as a function of trap timescale} \label{sec:fourpanels}
The impact of charge trapping is dependent on the defect responsible. 
Figure~\ref{fig:raw4panel} demonstrates the effect of charge trap 
species with different release times $\tau$ on various scientific observables.
To compute each datum (filled symbols), 
we simulate $10^{7}$ galaxies, add shot noise, add CTI
trailing in the $y$ direction (i.e.\ vertical in 
Fig.~\ref{fig:sampleimage}), only then add readout noise. 
Separately, we simulate $10^{6}$ stars. 
Using eqs.~\eqref{eq:estim}--\eqref{eq:formal_e2}, we measure mean values
of photometry (top panel), astrometry (second panel) and morphology 
(size in the third, and ellipticity in the bottom panel). 
Our results confirm what \citet{2010PASP..122..439R} found in a different context. 

Three trap regimes are apparent, for all observables.
Very fast traps ($\tau\!\la\!0.3$\,transfers) do not displace
electrons far from the object; thus their effect on photometry is minimal
(top plot in Fig.~\ref{fig:raw4panel}). 
We observe significant relative changes in position,
size, and ellipticity, forming a plateau at low $\tau$, because 
even if captured electrons are released after the shortest amount of time, 
some of them will be counted one pixel off their origin. 
This is probably an artifact: We expect the effect of traps with $\tau\!<\!0.1$ 
to be different in an model simulating the transfer between the constituent
electrodes of the physical pixels, rather than entire pixels.

Very slow traps ($\tau\!\ga\!30$\,transfers) result in electrons being carried away over a long distance
such that they can no longer be assigned to their original source image.
Hence, they cause a charge loss compared to the CTI-free case.
However, because charge is removed from nearly everywhere in the image, their impact
on astrometry and morphology is small.

The most interesting behaviour is seen in the transitional region, for traps 
with a characteristic release time of a few transfer times. 
If electrons re-emerge several pixels from their origin, they are close enough 
to be still associated with their source image, but yield the strongest
distortions in size and ellipticity measurements. 
This produces local maxima in the lower two panels of Fig.~\ref{fig:raw4panel}.
If these measurements are scientifically important, performance can 
-- to some degree -- be optimised by adjusting a CCD's clock 
speed or operating temperature to move release times
outside the most critical range $1\!\la\!\tau\!\la\!10$
\citep{2012SPIE.8453E..17M}. 

In the star simulations (crosses in Fig.~\ref{fig:raw4panel} for degraded
images, plus signs for CTI-corrected images), the CTI effects are generally
smaller than for the faint galaxies, because the stars we simulate are brighter
and thus experience less trailing \emph{relative to their signal}.
Still, we measure about the same spurious ellipticity $\Delta e_{1}$ and even
a slightly higher relative size bias $\Delta R^{2}/R^{2}_{\mathrm{true}}$ for
the stars. The explanation is that the quadratic terms in the second-order
moments (eq.~\ref{eq:qmom}) allow for larger contributions from the outskirts 
of the object, given the right circumstances. 
In particular, the wider window size $w$ explains the differences between the
galaxy and PSF simulations. Notably, the peak in the 
$\Delta e_{1}(\tau)$ and
$\Delta R^{2}/R^{2}_{\mathrm{true}}(\tau)$ curves shifts from 
$\sim\!3\,\mbox{px}$ for the galaxies to $\sim\!9\,\mbox{px}$ for the stars.
Because the wider window function gives more weight to pixels away
from the centroid, the photometry becomes more sensitive to slower traps.

For a limited number of trap configurations, we have also tried varying the
trap density or the number of transfers (i.e.\ object position on the CCD). 
In both cases, the dependence is linear.
Overall, for all tested observables, the measurements in
the degraded images (Fig.~\ref{fig:raw4panel}, solid symbols) 
are well-fit by the empirical fitting function
\begin{multline}
f^{\mathrm{degrade}}(\rho,\tau)=\rho\Big(A+D_{\mathrm{a}}\,{\mathrm{atan}}{((\log{\tau}-D_{\mathrm{p}})/D_{\mathrm{w}})}+\\
G_{\mathrm{a}}\exp{((\log{\tau}-G_{\mathrm{p}})^2/2G_{\mathrm{w}}^2)}\Big),
\label{eq:adg}
\end{multline}
which combines an arc-tangent drop (``D'') and a Gaussian peak (``G'').
The best fit-amplitudes ($A$, $D_{\mathrm{a}}$ and $G_{\mathrm{a}}$), positions
on the $\tau$ axis ($D_{\mathrm{p}}$ and $G_{\mathrm{p}}$) and widths
($D_{\mathrm{w}}$ and $G_{\mathrm{w}}$), are listed in Table~\ref{tab:taufits}. 
The same functional form provides a good match to the residuals after CTI 
correction, $f^{\mathrm{resid}}(\rho,\tau)$ (open symbols in 
Fig.~\ref{fig:raw4panel}). These residuals are caused by readout noise,
which is not subject to CTI trailing, but undergoes CTI correction 
(see Sect.~\ref{sec:rn}).

\subsection{Predictive model for imperfect correction}

We set out to construct a predictive model $\Delta f^{\mathrm{Pr}}$
of $\Delta\eta$, the CTI effect in an observable relative to the underlying
true value (eq.~\ref{eq:deltadef}). There are two terms, the CTI degradation
(eq.~\ref{eq:adg}), and a second term for the effect of the 
`inverse' CTI correction allowing for a slightly imperfect CTI model:
\begin{equation}
\Delta f^\mathrm{Pr}=f^\mathrm{degr}(\rho,\tau)+f^\mathrm{correct}(\rho+\Delta\rho,\tau+\Delta\tau).
\label{eq:adg2}
\end{equation}
Since CTI trailing perturbs an image by only a small amount, the correction 
acts on an almost identical image. 
Assuming the coefficients of eq.~(\ref{eq:adg}) to be constant, we get:
\begin{equation} \label{eq:prediction}
\Delta f^\mathrm{Pr}\approx f^\mathrm{degr}(\rho,\tau)-f^\mathrm{degr}(\rho+\Delta\rho,\tau+\Delta\tau) 
+ f^{\mathrm{res}}(\rho,\tau), 
\end{equation}
where $f^{\mathrm{res}}(\rho,\tau)$ is approximately constant, and depends 
on the readout noise (see Section~\ref{sec:zprn}). We could expand this equation 
as a Taylor series, but the derivatives of $f$ do not provide much further insight.

Because eq.~(\ref{eq:nenc}) is non-linear in the number $n_{\mathrm{e}}$ of 
signal electrons, 
our observation (Sect.~\ref{sec:fourpanels}) that the 
\emph{effects} of CTI behave linearly in $\rho$ is not a trivial result.
Assuming this linearly in $\rho$, we can expand
eq.~(\ref{eq:prediction}) and factor out $\rho$. The combined effect of
several trap species $i$ with release timescales $\tau_{i}$ and densities 
$\rho_{i}$ can then be written as:
\begin{multline} \label{eq:sumpred}
\Delta f^{\mathrm{Pr}}(\rho_{i}+\Delta\rho_{i},\tau_{i}+\Delta\tau_{i})\!
=\!\sum_{i}\rho_{i}f^{\mathrm{resid}}(\tau_{i}) + \\ 
\sum_{i}\left[\rho_{i}f(\tau_{i}) - (\rho_{i}+\Delta\rho_{i})f(\tau_{i}+\Delta\tau_{i})\right],
\end{multline}
in which we dropped the superscript of $f^{\mathrm{degr}}$ for the sake
of legibility. We are going to test this model in the remainder of this study,
where we consider a mixture of three trap species. We find eq.~(\ref{eq:sumpred})
to correctly describe measurements of spurious ellipticity $\Delta e_{1}$, as
well as the relative bias in source size $\Delta R^{2}/R^{2}_{\mathrm{true}}$
and flux $\Delta F/F_{\mathrm{true}}$.

\section{Euclid as a concrete example} \label{sec:euclid}

\subsection{Context for this study}

To test the general prediction eq.~(\ref{eq:sumpred}), we now evaluate the 
effect of imperfect CTI correction in simulations of {\it Euclid} data, 
with a full {\it Euclid} CTI model featuring multiple trap species (see Sect.~\ref{sec:blm}).
We call this the $\mathbf{M}+\delta\mathbf{M}$ experiment. 

Akin to \citet{2012MNRAS.419.2995P} for \textit{Gaia}, this
study is useful in the larger context of the flow down of requirements from 
\textit{Euclid}'s science goals
\citep{2010arXiv1001.0061R} to its imaging capabilities
\citep{2013MNRAS.429..661M} and instrument implementation
\citep{2013MNRAS.431.3103C,2014SPIE.9143E..0JC}.
In particular, \citet{2013MNRAS.429..661M} highlight that the mission's overall 
success will be determined both by its absolute instrumental
performance and our knowledge about it. We now present 
the next step in the flow down:
to what accuracy do we need to constrain the parameters of the
\citet{2014MNRAS.439..887M} CTI model? 
Future work will then determine which
calibration observations are required to achieve this accuracy.

While the final \textit{Euclid} requirements remain to be confirmed,
we adopt the current values as discussed by \citet{2013MNRAS.431.3103C}.
Foremost, the ``CTI contribution to the PSF 
ellipticity shall be $<\!1.1\times10^{-4}$ per ellipticity component''. 

The \textit{Euclid} VIS PSF model will bear an uncertainty due to CTI,
that translates into an additional error on measured galaxy properties.
For the bright stars (which have much higher $S/N$) tracing the PSF,
\citet{2013MNRAS.431.3103C} quote a
required knowledge of $R^{2}$ to a precision 
$\left|\sigma(R^{2})\right|\!<\!4\times10^{-4}$.
We test this requirement with our second suite of simulations,
containing $10^{6}$ realisations of a \textit{Euclid} VIS PSF at
$S/N\!\approx\!200$ (cf.~Sec.~\ref{sec:imsim}).
 
In reality, CTI affects the charge transport in both CCD directions,
serial and parallel. For the sake of simplicity, we only consider serial CTI,
and thus underestimate the total charge trailing.
There is no explicit photometric error budget allocated to CTI, while
``ground data processing shall correct 
for the detection chain response to better than $0.7$~\% error in photometry 
in the nominal VIS science images''.

\subsection{CTI model for the \textit{Euclid} VISual instrument} \label{sec:blm}

\begin{table}
\caption{The baseline trap model $\mathcal{M}$. 
The model includes a baseline well fill power of $\beta_{0}\!=\!0.58$.}
\begin{tabular}{cccc}\hline\hline
\textbf{Baseline model} & $i\!=\!1$ & $i\!=\!2$ & $i\!=\!3$ \\
Trap density $\rho_{i}$ [px$^{-1}$] & $0.02$ & $0.03$ & $0.95$ \\
Release timescale $\tau_{i}$ [px] & $0.8$ & $3.5$ & $20.0$ \\
\hline\hline \label{tab:traps}
\end{tabular}
\end{table}
\begin{figure*}
\includegraphics[width=88.1mm,angle=180]{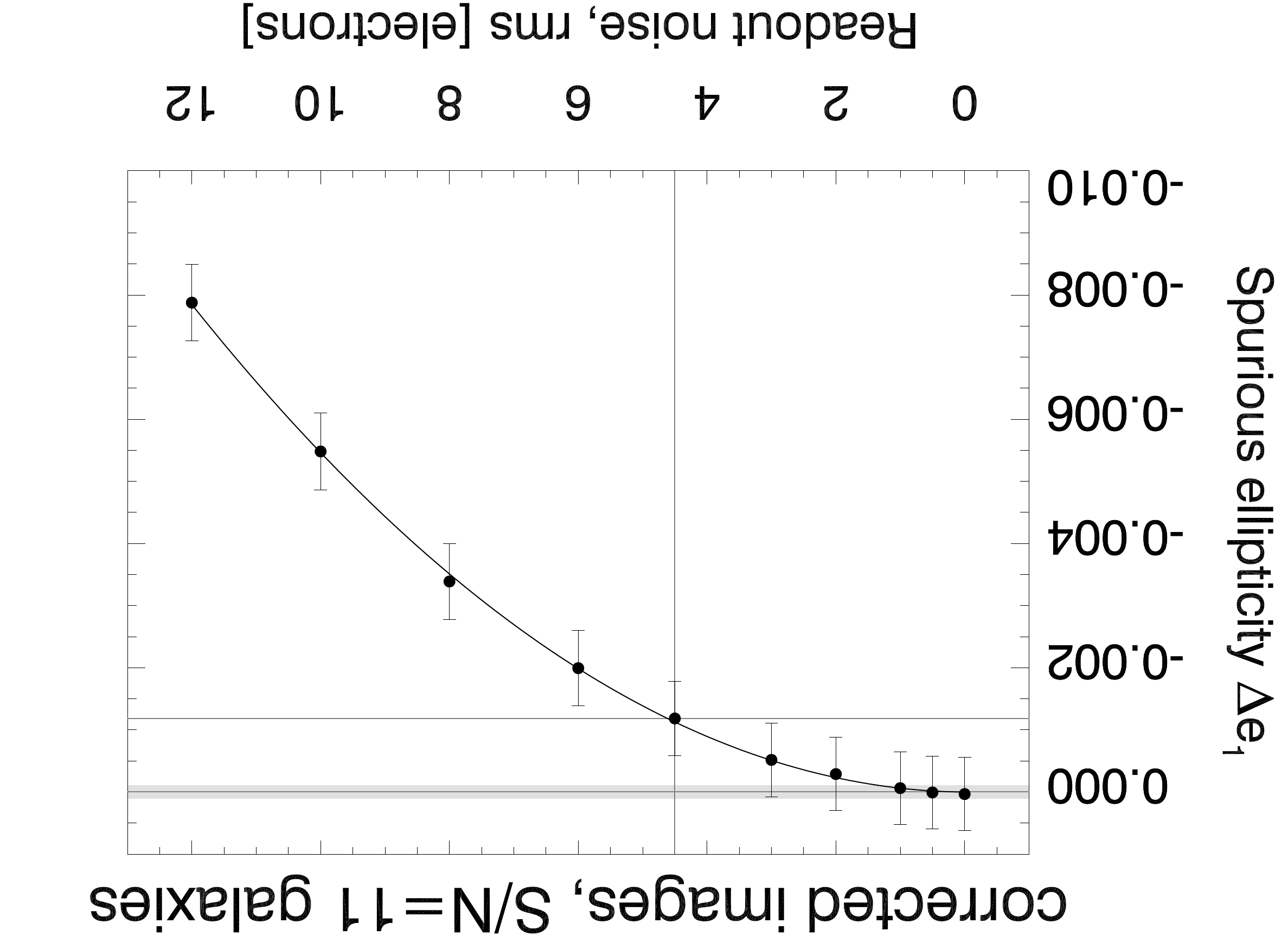}
\includegraphics[width=88.1mm,angle=180]{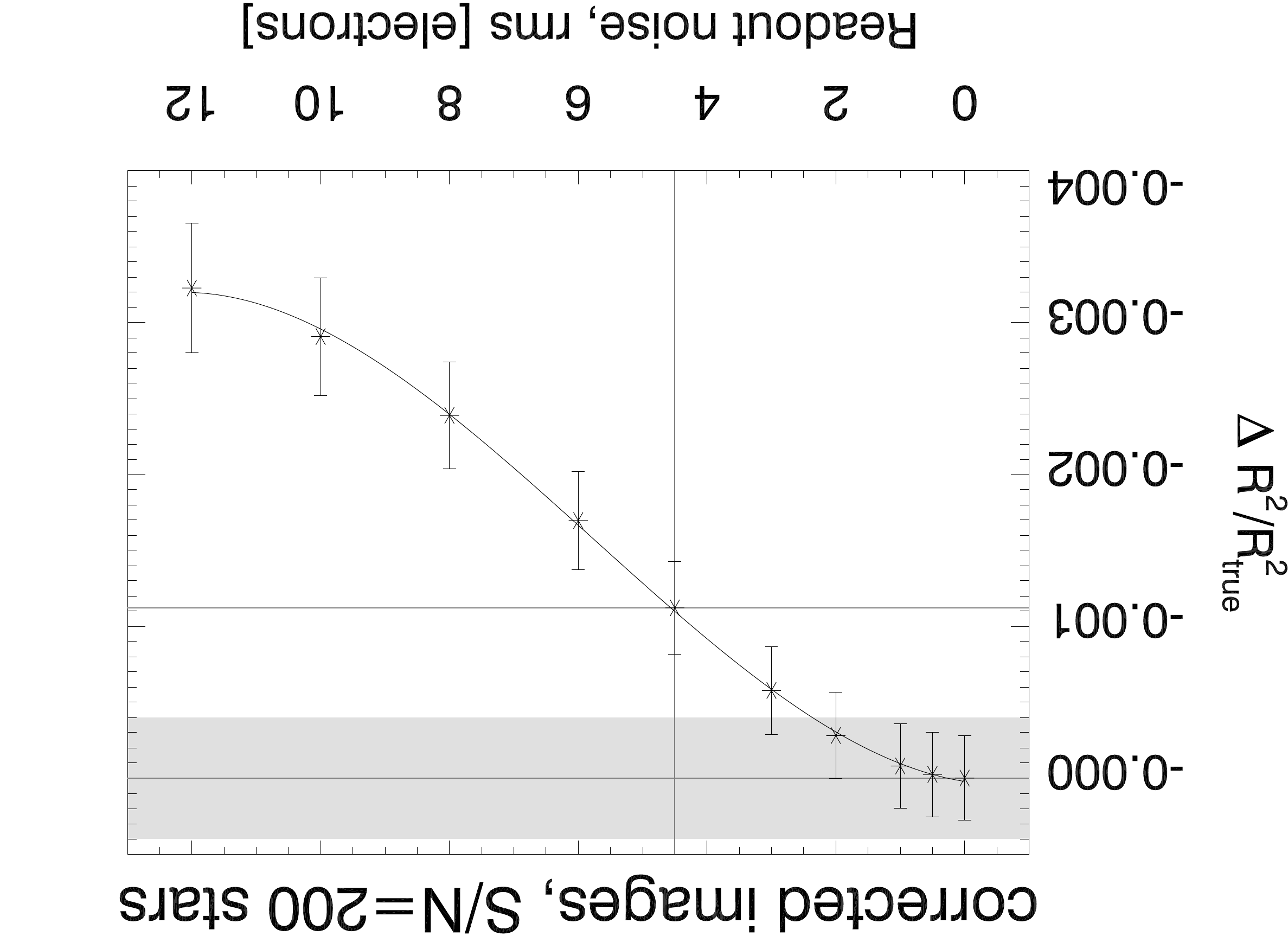}
\caption{Residual spurious ellipticity and source size induced by CTI  
after correction, as a function of the rms readout noise. 
\textit{Left plot:} Spurious ellipticity $\Delta e_{1}$ in the faintest 
galaxies to be analysed for \textit{Euclid}, at the end of the mission,
and furthest from the readout register.
\textit{Right plot:} Bias in source size $\Delta R^{2}/R^{2}_{\mathrm{true}}$ 
in bright stars, at the end of the mission, and furthest from the readout register.
Lines show the best quadratic (cubic) fits.
Shaded regions indicate the \textit{Euclid} VIS requirements.
Vertical grey lines mark the nominal rms readout noise of $4.5$ electrons.}
\label{fig:readnoise}
\end{figure*}
Based on a suite of laboratory test data, we define a baseline model 
$\mathbf{M}$ of the most important CTI parameters ($\rho_{i}$, $\tau_{i}$, $\beta_{0}$).
We degrade our set of $10^{7}$ simulated galaxies using $\mathbf{M}$.
The $\mathbf{M}+\delta\mathbf{M}$ experiment then consists of correcting
the trailing in the degraded images with slight alterations to $\mathbf{M}$.
We investigate $>\!100$ correction models $\mathbf{M}+\delta\mathbf{M}$,
resulting in an impressive $1.4\times10^{9}$ simulated galaxies used in this study.

Exposure to the radiation environment in space was simulated in the laboratory
by irradiating a prototype of the e2v CCD273 to be used for \textit{Euclid} VIS
with a $10$~MeV equivalent fluence of $4.8\!\times\!10^{9}\,\mathrm{protons/cm}^{-2}$
\citep{2014P1P,2014V1V}. 
Characterisation experiments were performed in nominal VIS conditions of 
$153\,\mbox{K}$ temperature and a $70\,\mbox{kHz}$ readout frequency.
We refer to Appendix~\ref{sec:labdata} for further details on the experiments and data analysis.

We emphasize that our results for $e_{1}$
pertain to faint and small galaxies, with 
an exponential disk profile (vz.~Sect.~\ref{sec:imsim}), and placed at the
maximum distance from the readout register ($y\!=\!2051$ transfers). 
Furthermore, we assume the level of radiation damage expected at the end of
\textit{Euclid}'s six year mission.
Because CTI trailing increases roughly linearly with time in orbit
\citep[cf.][]{2014MNRAS.439..887M}, the CTI experienced by the typical
faintest galaxy (i.e.\ at half the maximum distance 
to the readout register and three 
years into the mission), will be smaller by a factor of $4$
compared to the results quoted below. 

Where not stated otherwise the nominal \textit{Euclid} VIS rms readout noise 
of $4.5$ electrons was used.
Table~\ref{tab:traps} summarises the baseline model $\mathbf{M}$ that was 
constructed based on these analyses. The default well fill power is 
$\beta_{0}\!=\!0.58$. Slow traps with $\tau_{3}\!=\!20$ clock
cycles and $\rho_{3}\!=\!0.95$ dominate our baseline model, with small 
fractions of medium-fast ($\tau_{2}\!=\!3.5$, $\rho_{2}\!=\!0.03$) and fast 
($\tau_{1}\!=\!0.8$, $\rho_{1}\!=\!0.02$) traps.
Figure~\ref{fig:models} shows how trails change with changing trap parameters.

\subsection{Readout noise impedes perfect CTI correction} \label{sec:zprn}

\subsubsection{Not quite there yet: the zeropoint} \label{sec:zp}

First, we consider the ellipticities measured in the degraded and corrected
images, applying the same baseline model in the degradation and correction
steps. The reasons why this
experiment does not retrieve the same corrected ellipticity $e_{\mathrm{corr}}$
as input ellipticity $e_{\mathrm{in}}$ are the
Poissonian image noise and Gaussian readout noise. We quantify this in terms of
spurious ellipticity $\Delta e\!=\!e_{\mathrm{corr}}-e_{\mathrm{in}}$, 
and shall refer to it as the \textit{zeropoint} of the $\mathbf{M}+\delta\mathbf{M}$
experiment. 
The spurious ellipticity in the serial direction is 
$Z_{\mathrm{e_{1}}}\!=\!\Delta e_{1}\!=\!-0.00118\pm0.00060$. 
Thus, this experiment on worst-case galaxies using the current software 
exceeds the   
\textit{Euclid} requirement of $\left|\Delta e_{\alpha}\right|\!<\!1.1\times10^{-4}$ 
by a factor of $\sim\!10$.
With respect to the degraded image $99.68$~\% of the CTI-induced ellipticity 
are being corrected. Virtually the same 
zeropoint, $\Delta e_{1}\!=\!-0.00118\pm0.00058$, is predicted 
by adding the contributions of the three species
from single-species runs based on the full $10^{7}$ galaxies. 
We point out that these results on the faintest galaxies furthest from the 
readout register have been obtained using non-flight 
readout electronics \citep[cf.][]{2014SPIE.9154E..0RS}. 

From our simulation of $10^{6}$ bright ($S/N\!\approx\!200$) stars,
we measure the residual bias in source size $R^{2}$ after CTI correction of
$Z_{\!R^{2}}\!=\!\Delta R^{2}/R^{2}_{\mathrm{true}}\!=\!(-0.00112\pm0.00030)$,
in moderate excess of the requirement 
$\left|\Delta R^{2}/R^{2}_{\mathrm{true}}\right|\!<\!4\times10^{-4}$. 
While the $S/N$ of the star simulations is selected to represent the typical
\textit{Euclid} VIS PSF tracers, the same arguments of longest distance from
the readout register and non-flight status of the electronics apply.

\subsubsection{The effect of readout noise} \label{sec:rn}

In Fig.~\ref{fig:readnoise}, we explore the effect of varying the rms readout noise in
our simulations about the nominal value of $4.5$ electrons (grey lines)
discussed in Sect.~\ref{sec:zp}. We continue to use the baseline
trap model for both degradation and correction. 
For the rms readout noise, a range of values between $0$ and $15$ 
electrons was assumed. 
For the faint galaxies (Fig.~\ref{fig:readnoise}, left plot), we find
$\Delta e_{1}$ to increase with readout noise in a way well described
by a second-order polynomial. A similar, cubic fit can be found for 
$\Delta R^{2}/R^{2}_{\mathrm{true}}$ measured from the star simulations
(Fig.~\ref{fig:readnoise}, right plot), but with a hint towards saturation
in the highest tested readout noise level. 

The most important result from Fig.~\ref{fig:readnoise} is that in absence of readout noise,
if the correction assumes the correct trap model $\mathbf{M}$, it 
removes the CTI perfectly, with $\Delta e_{1}\!=\!(0.3\pm5.9)\times 10^{-4}$ and
$\Delta R^{2}/R^{2}_{\mathrm{true}}\!=\!(0.0\pm2.8)\times 10^{-4}$. 
The quoted uncertainties are determined by the $N\!=\!10^{7}$ 
($10^{6}$) galaxy images 
we simulated. We conclude that the combination 
of our simulations and correction code pass this crucial sanity check. 
If the rms readout noise is $\lesssim\!3$ electrons 
($\lesssim\!0.5$ electrons), the spurious ellipticity (the relative size bias) 
stays within \textit{Euclid} requirements.

\subsection{Sensitivity to imperfect CTI modelling} \label{sec:res}

\subsubsection{Morphology biases as a function of well fill power,
and determining tolerance ranges} \label{sec:beta}

\begin{figure}
\includegraphics[width=85mm,angle=180]{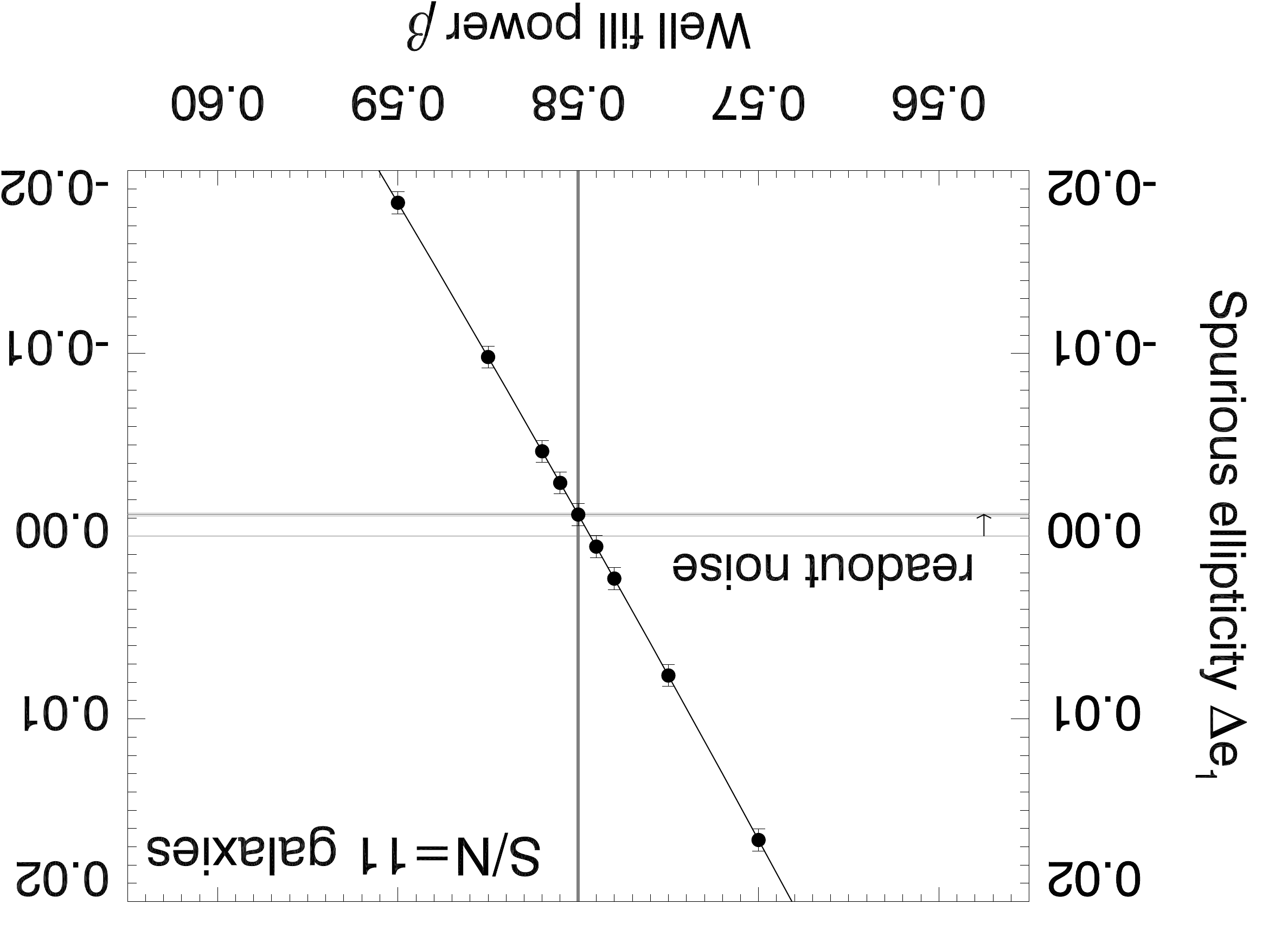}
\includegraphics[width=85mm,angle=180]{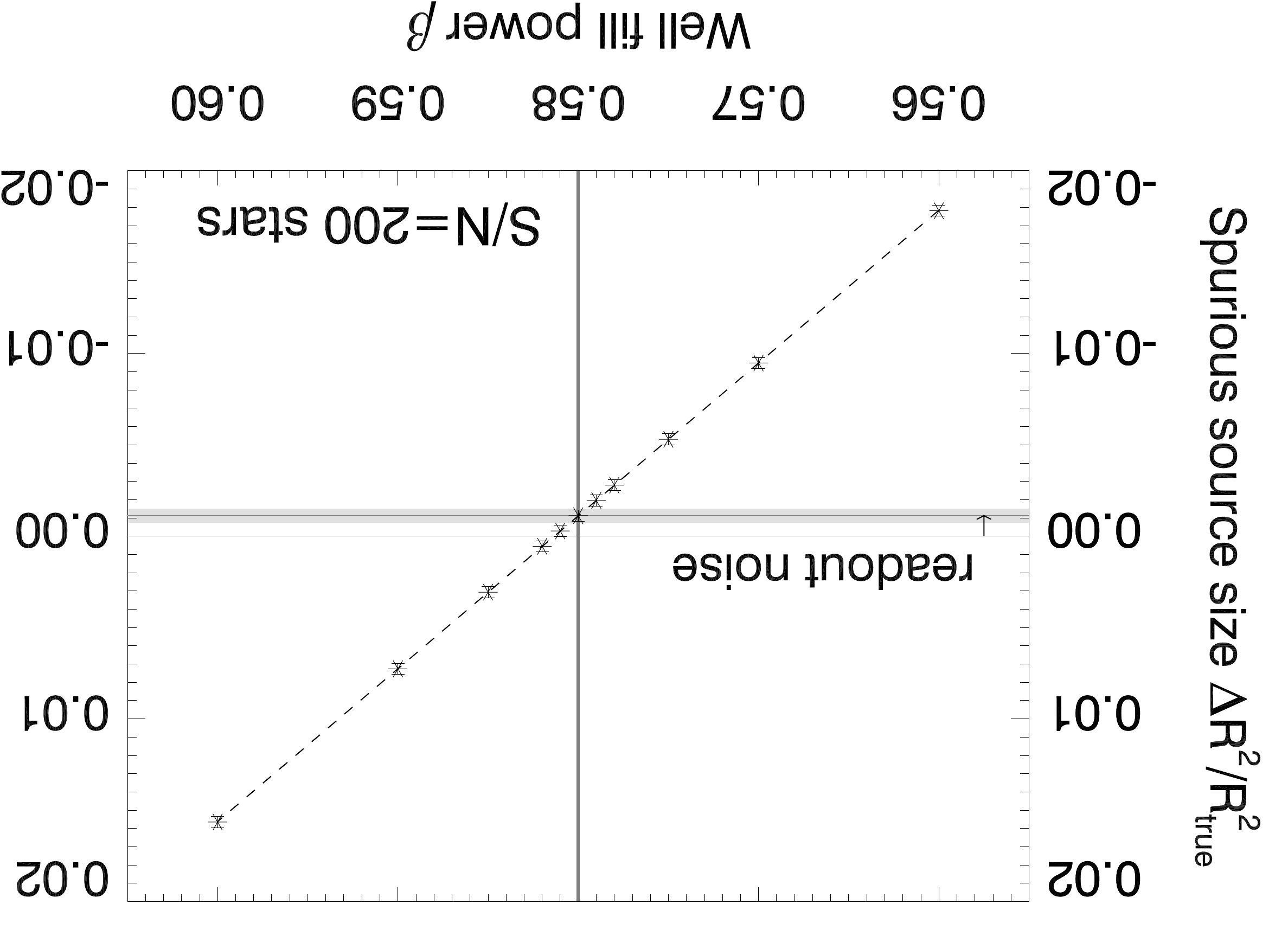}
\caption{Sensitivity of the CTI-induced spurious ellipticity 
$\Delta e_{1}$ \textit{(upper plot)} and the relative spurious source size 
$\Delta R^{2}/R^{2}_{\mathrm{true}}$ \textit{(lower plot)} to the well fill power $\beta$. 
At the default value of $\beta\!=\!0.58$ (vertical grey line), the 
measurements deviate from zero due to readout noise, 
as indicated by arrows. The shaded region around
the measurements indicate the \textit{Euclid} 
requirement ranges as a visual aid.
Solid and dashed lines display quadratic (linear) 
fits to the measured $\Delta e_{1}(\beta)$ and 
$\Delta R^{2}(\beta)/R^{2}_{\mathrm{true}}$, respectively. 
We study the worst affected objects (at the end of the mission
and furthest from the readout register) and the faintest \textit{Euclid}
galaxies. This plot also assumes CTI is calibrated from charge injection
lines at full well capacity only. This will not be the case.}
\label{fig:mdm2_exp0}
\end{figure}

Now that we have assessed the performance of the correction using 
the same CTI model as for the degradation (given the specifications of our
simulations), we turn to the $\mathbf{M}+\delta\mathbf{M}$ experiment for
determining the sensitivities to imperfections in the CTI model.
To this end, we assume the zeropoint offset $Z_{e_{1}}$ (or $Z_{\!R^{2}}$) 
of Sect.~\ref{sec:zp} to be corrected, and `shift' the requirement range 
to be centred on it (see, e.g., Fig.~\ref{fig:mdm2_exp0}).

Figure~\ref{fig:mdm2_exp0} 
shows the $\mathbf{M}+\delta\mathbf{M}$ experiment
for the well fill power $\beta$. If the degraded images are corrected with the
baseline $\beta_{0}\!=\!0.58$, we retrieve the zeropoint measurement from 
Sect.~\ref{sec:zp}. For the $\mathbf{M}+\delta\mathbf{M}$ experiment, we
corrected the degraded images with slightly different well fill powers
$0.56\!\leq\!\beta\!\leq\!0.60$. The upper plot in 
Fig.~\ref{fig:mdm2_exp0} shows the resulting $\Delta e_{1}$ 
in galaxies, and the lower plot 
$\Delta R^{2}/R^{2}_{\mathrm{true}}$ in stars. 
We find a strong dependence of both
the spurious serial ellipticity $\Delta e_{1}$
and $\Delta R^{2}/R^{2}_{\mathrm{true}}$ on $\Delta\beta\!=\!\beta\!-\!\beta_{0}$.

In order to determine a tolerance range with respect to a CTI model parameter
$\xi$ with baseline value $\xi_{0}$ (here, the well fill power $\beta$), we fit 
the measured bias $\Delta\eta$ (e.g.\ $\Delta e_{1}$, cf.\ eq.~\ref{eq:deltadef})
as a function of $\Delta\xi\!=\!\xi\!-\!\xi_{0}$. By assuming a polynomial
\begin{equation} \label{eq:polynom}
\Delta\eta(\Delta\xi) = Z_{\eta} + \sum_{j=1}^{J}{a_{j}(\Delta\xi)^{j}}
\end{equation}
of low order $J$, we perform a Taylor expansion around $\xi_{0}$.
In eq.~\ref{eq:polynom}, $Z_{\eta}$ is the zeropoint (Sect.~\ref{sec:zp})
to which we have shifted our requirement margin.
The coefficients $a_{j}$ are determined using 
the \texttt{IDL} singular value decomposition least-square fitting routine \texttt{SVDFIT}. 
For consistency, our fits include $Z_{\eta}$ as the zeroth order.
In Fig.~\ref{fig:mdm2_exp0}, the best-fitting quadratic 
(linear) fits to $\Delta e_{1}$ ($\Delta R^{2}/R^{2}_{\mathrm{true}}$) 
are shown as a solid and dashed line, respectively.

In both plots, the data stick tightly to the best-fitting lines, 
given the measurement uncertainties.
If the measurements were uncorrelated, this would be a suspiciously orderly trend.
However, as already pointed out in Sect.~\ref{sec:sigmas}, we re-use the same
$10^{7}$ simulations with the same peaks and troughs in the noise in all data
points shown in Figs.~\ref{fig:mdm2_exp0} to \ref{fig:flux2}.
Hence, we do not expect to see data points to deviate from the regression lines
to the degree their quoted uncertainties would indicate. As a consequence,
we do not make use of the $\chi^{2}_{\mathrm{red}}\!\ll\!1$ our fits commonly
yield for any interpretation.

Because the interpretation of the reduced $\chi^{2}$ is tainted by the 
correlation between our data points, we use an alternative criterion to decide
the degree $J$ of the polynomial: If the uncertainty returned by 
\texttt{SVDFIT} allows for a coefficient $a_{j}\!=\!0$, 
we do not consider this or higher terms. 
For the panels of Fig.~\ref{fig:mdm2_exp0}, this procedure yields $J\!=\!2$ ($J\!=\!1$). 
The different signs of the slopes are expected because $R^{2}$ appears in the
denominator of eq.~(\ref{eq:chidef}).

Given a requirement $\Delta\eta_{\mathrm{req}}$, e.g. 
$\Delta e_{1,\mathrm{req}}\!=\!1.1\times10^{-4}$,
the parametric form (eq.~\ref{eq:polynom}) of the sensitivity curves allows
us to derive tolerance ranges to changes in the trap parameters.
Assuming the zeropoint (the bias at the correct value of $\xi$) to be 
accounted for, we find the limits of the tolerance range as the solutions 
$\Delta\xi_{\mathrm{tol}}$ of
\begin{equation} \label{eq:tol}
\left|\sum_{j=1}^{J}{a_{j}(\Delta\xi)^{j}}\right|\!=\!\Delta\eta_{\mathrm{req}}
\end{equation}
with the smallest values of $|\Delta\xi|$ on either sides to $\Delta\xi\!=\!0$.
Using, eq.~(\ref{eq:tol}), we obtain 
$\Delta\beta_{\mathrm{tol}}\!=\!\pm(6.31\pm0.07)\times10^{-5}$ 
from the requirement on the spurious ellipticity $\Delta e_{1}\!<\!1.1\times10^{-4}$, 
for which the quadratic term is small. From the requirement on the relative 
size bias $\Delta R^{2}/R^{2}_{\mathrm{true}}\!<\!4\!\times\!10^{-4}$ 
we obtain $\Delta\beta_{\mathrm{tol}}\!=\!\pm(4.78\pm0.05)\times10^{-4}$. 
In other words, the ellipticity sets the more stringent requirement, and 
we need to be able to constrain $\beta$ to an accuracy of at least 
$(6.31\pm0.07)\times10^{-5}$ in absolute terms. 
This analysis assumes calibration by a single charge injection line at full well
capacity, such that eq.~(\ref{eq:nenc}) 
needs to be extrapolated to lower signal levels. 
We acknowledge that \textit{Euclid} planning has already adopted 
using also faint charge injection lines, lessening the need to extrapolate.

\subsubsection{Ellipticity bias as a function of trap density} \label{sec:rho}
\begin{figure*}
\includegraphics[width=140mm,angle=180]{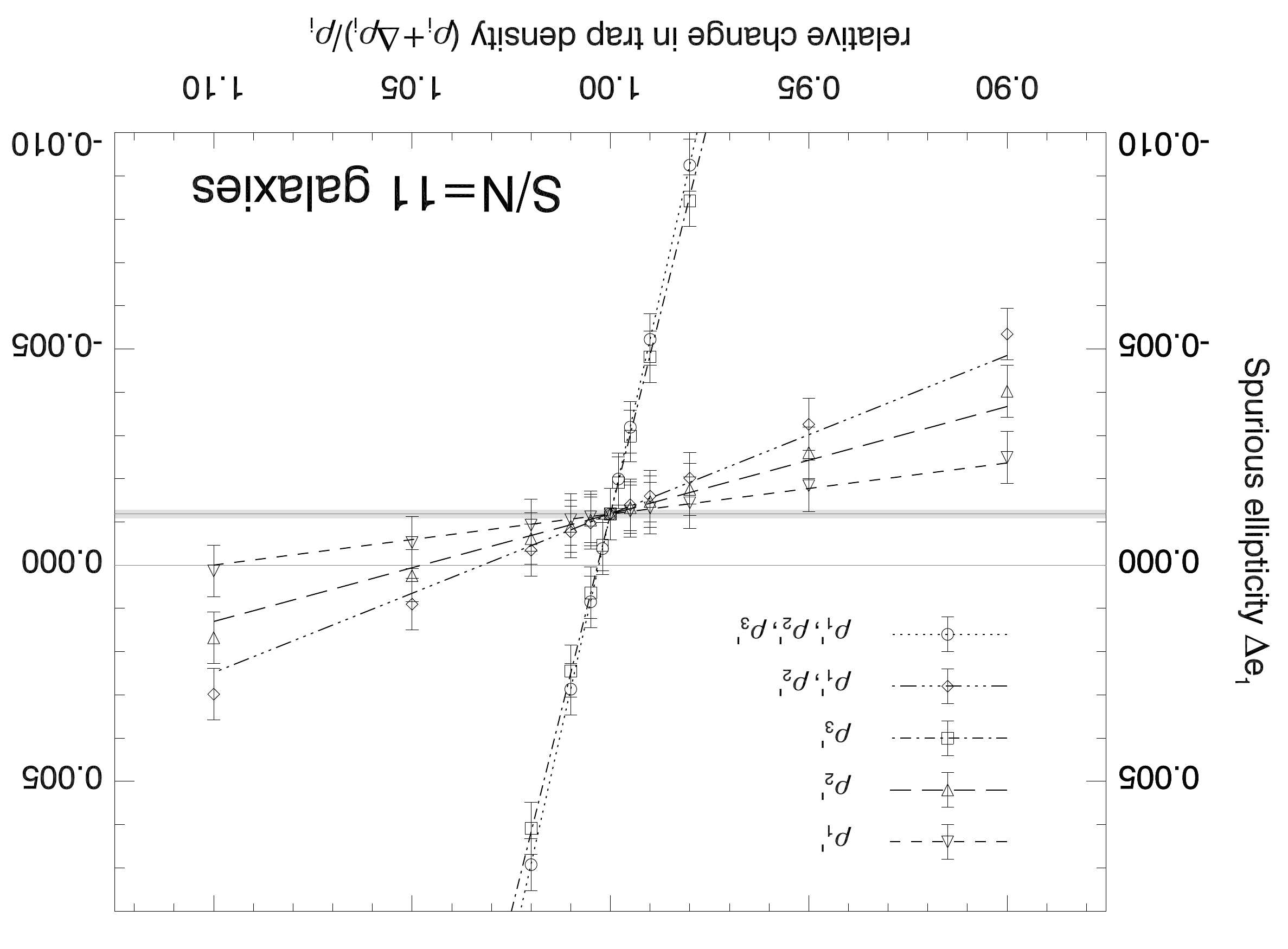}
\includegraphics[width=140mm,angle=180]{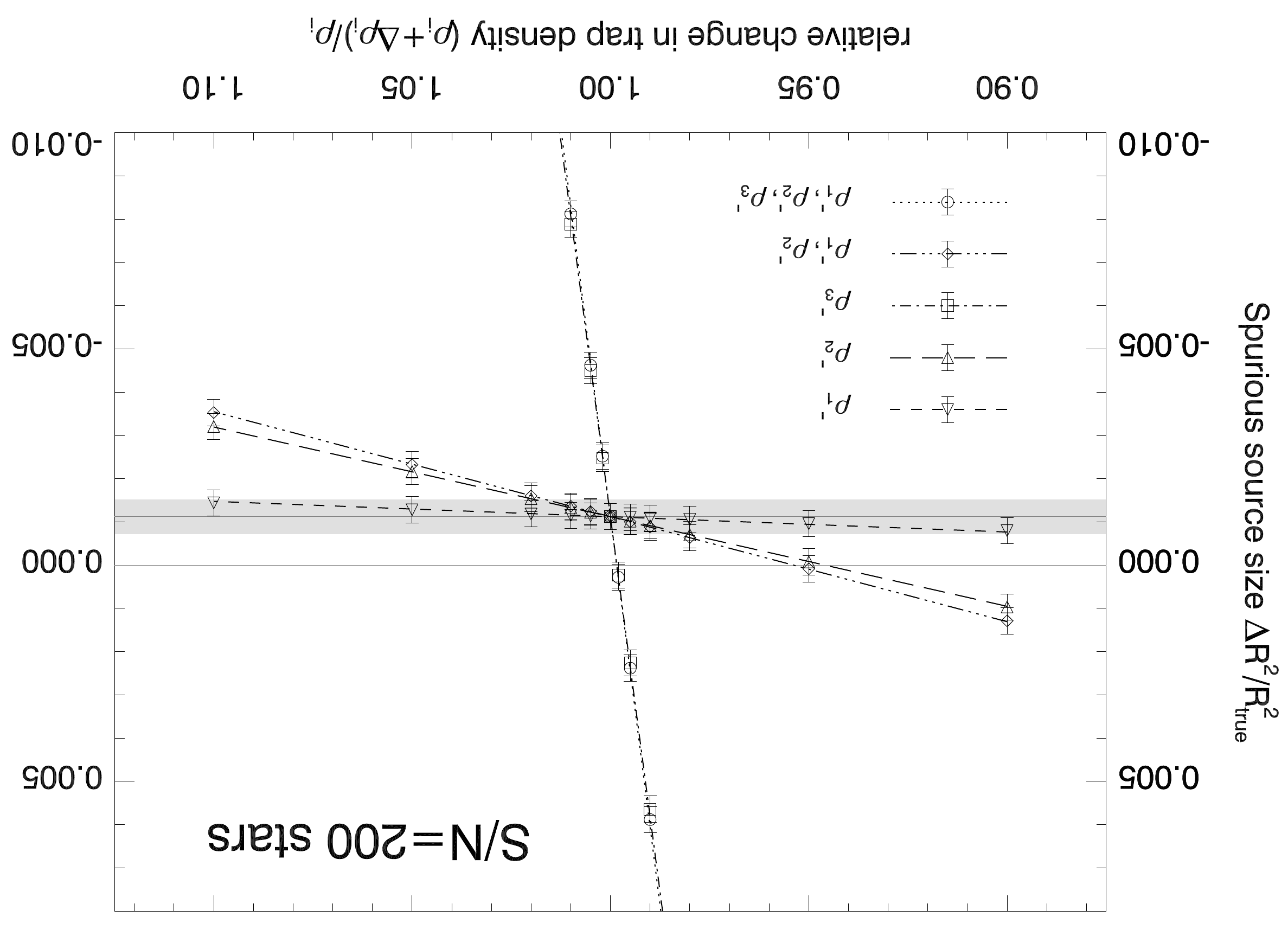}
\caption{Sensitivity of the CTI-induced spurious ellipticity $\Delta e_{1}$ 
in faint galaxies 
\textit{(upper panel)} and relative bias in source size 
$\Delta R^{2}/R^{2}_{\mathrm{true}}$ \textit{(lower panel)} 
in bright stars to a 
relative change in trap densities $(\rho_{i}+\Delta\rho_{i})/\rho_{i}$. Different 
symbols and line styles denote which to which of the trap species a change in density 
was applied: The slow traps: ($\rho_{1}$, upward and dashed line); the medium traps: 
($\rho_{2}$, downward triangles and long-dashed line); both of them: ($\rho_{1}$, $\rho_{2}$,
diamonds and triple dot-dashed line); the fast traps ($\rho_{3}$,
squares and dot-dashed line); all: ($\rho_{1}$, $\rho_{2}$, $\rho_{3}$,
circles and dotted line). 
The various broken lines show the best-fit representation of the
measurements as given by the sensitivity model (Eq.~\ref{eq:rhopred}). 
Like in Fig.~\ref{fig:mdm2_exp0}, the grey shaded area
indicates the \textit{Euclid} VIS requirement range. 
We study the worst affected objects (at the end of the mission
and furthest from the readout register) and the faintest \textit{Euclid}
galaxies.}
\label{fig:mdm2_exp1}
\end{figure*}
We now analyse the sensitivity of $\Delta e_{\alpha}$ towards changes in the
trap densities.
Figure~\ref{fig:mdm2_exp1} shows the $\mathbf{M}+\delta\mathbf{M}$ experiment
for one or more of the trap densities $\rho_{i}$ of the baseline model.
The upper panel of Fig.~\ref{fig:mdm2_exp1} presents the spurious ellipticity
$\Delta e_{1}$ for five different branches of the experiment.
In each of the branches, we modify the densities $\rho_{i}$ of one or several 
of the trap species. For example, the upward triangles 
in Fig.~\ref{fig:mdm2_exp1} denote that the correction model applied to the 
images degraded with the baseline model used a density of 
the fast trap species $\rho_{1}\!+\!\Delta\rho_{1}$, tested at several values 
of $\Delta\rho_{1}$ with $0.9\!\leq\!1\!+\!\Delta\rho_{1}/\rho_{1}\!\leq\!1.1$. 
The densities of the other species are kept to their baseline values in this case. 
The other four branches modify $\rho_{2}$ (downward triangles); 
$\rho_{3}$ (squares); $\rho_{1}$ and $\rho_{2}$ (diamonds); and all 
three trap species (circles).

Because a value of $\Delta\rho_{i}\!\!=\!\!0$ 
reproduces the baseline model in all branches,
all of them recover the zeropoint measurement of $\Delta e_{1}$ there 
(cf.\ Sect.~\ref{sec:zp}).  
Noticing that $e_{\mathrm{degr,1}}\!-\!e_{\mathrm{in,1}}\!<\!0$ 
for the degraded images relative to the input images,
we explain the more negative
$\Delta e_{1}$ for $\Delta\rho_{i}\!<\!0$ as the effect of 
undercorrecting 
the CTI. This applies to all branches of the experiment. Likewise, with increasing 
$\Delta\rho_{i}\!>\!0$, the residual undercorrection at the zeropoint decreases.
Eventually, with even higher $\kappa\!>\!1$, we overcorrect the CTI and measure
$\Delta e_{1}\!>\!0$.

Over the range of $0.9\!\leq\!1\!+\!\Delta\rho_{1}/\rho_{1}\!\leq\!1.1$ 
we tested, $\Delta e_{1}$ responds 
linearly to a change in the densities. 
Indeed, our model (eq.~\ref{eq:sumpred}), which is linear in the $\rho_{i}$
and additive in the effects of the different trap species, provides an
excellent description of the measured data, both for $\Delta e_{1}$ and
$\Delta R^{2}/R^{2}_{\mathrm{true}}$ (Fig.~\ref{fig:mdm2_exp1}, lower panel).
The lines in Fig.~\ref{fig:mdm2_exp1} denote the model prediction from a
simplified version of eq.~(\ref{eq:sumpred}),
\begin{equation} \label{eq:rhopred}
\Delta f^{\mathrm{Pr}}(\rho_{i}+\Delta\rho_{i})\!=\!
\sum_{i}\rho_{i}f^{\mathrm{resid}}(\tau_{i}) + \sum_{i}(\rho_{i}+\Delta\rho_{i})f(\tau_{i})\,.
\end{equation}
In eq.~(\ref{eq:rhopred}), we assumed the $\tau_{i}$ be correct, 
i.e.\ $\Delta\tau_{i}\!=\!0$. 

Next, we compute the tolerance $\Delta\rho_{i,\mathrm{tol}}/\rho$ 
by which, for each branch of the experiment, we might deviate from the correct trap model
and still recover the zeropoint within the \textit{Euclid} requirements of
$\left|\Delta e_{\alpha,\mathrm{req}}\right|\!<\!1.1\times10^{-4}$, 
resp.~ $\left|\Delta R^{2}_{\mathrm{req}}/R^{2}_{\mathrm{true}}\right|\!<\!4\times10^{-4}$. 
Again, we calculate these tolerances about the zeropoints
$Z\!=\!\sum_{i}\rho_{i}f^{\mathrm{resid}}(\tau_{i})$ (cf.~eq.~\ref{eq:rhopred}),
that we found to exceed the requirements in Sect.~\ref{sec:zp}, but assume to
be corrected for in this experiment.

In accordance with the linearity in $\Delta\rho_{i}$, applying the Taylor expansion
recipe of Sect.~\ref{sec:beta}, we find the data in Fig.~\ref{fig:mdm2_exp1} 
to be well represented by first-order polynomials (eq.~\ref{eq:polynom}).
The results for $\Delta\rho_{i,\mathrm{tol}}/\rho$ we obtain from eq.~(\ref{eq:tol}) 
are summarised in Table~\ref{tab:tolerances}. For all species,  
the constraints from $\Delta e_{1}$ for faint galaxies are tighter than the ones from 
$\Delta R^{2}/\Delta R^{2}_{\mathrm{true}}$ for bright stars. 

Only considering the fast traps, $\rho_{1}$ can change by $0.84\pm0.33$\%  
and still be within \textit{Euclid} VIS requirements, 
\textit{given the measured zeropoint has been corrected for}.
While a tolerance of $0.39\pm0.06$\% 
is found for $\rho_{2}$, the slow traps put a much tighter tolerance of 
$0.0303\pm0.0007$\% on the density $\rho_{3}$. 
This is expected because
slow traps amount to $95$\% of all baseline model traps 
(Table~\ref{tab:traps}). Varying the
density of all trap species in unison, we measure a tolerance of 
$0.0272\pm0.0005$\%.

Computing the weighted mean of the $\Delta\tau\!=\!0$ intercepts in
Fig.~\ref{fig:mdm2_exp1}, we derive better constraints on the zeropoints: 
$Z_{\mathrm{e_{1}}}\!=\!\Delta e_{1}\!=\!-0.00117\pm0.00008$ for the faint galaxies,
and $Z_{\!R^{2}}\!=\!\Delta R^{2}/R^{2}_{\mathrm{true}}\!=\!-0.00112\pm0.00004$
for the bright stars.

\subsubsection{Ellipticity bias as a function of trap release time} \label{sec:tau}

\begin{figure*}
\includegraphics[width=140mm,angle=180]{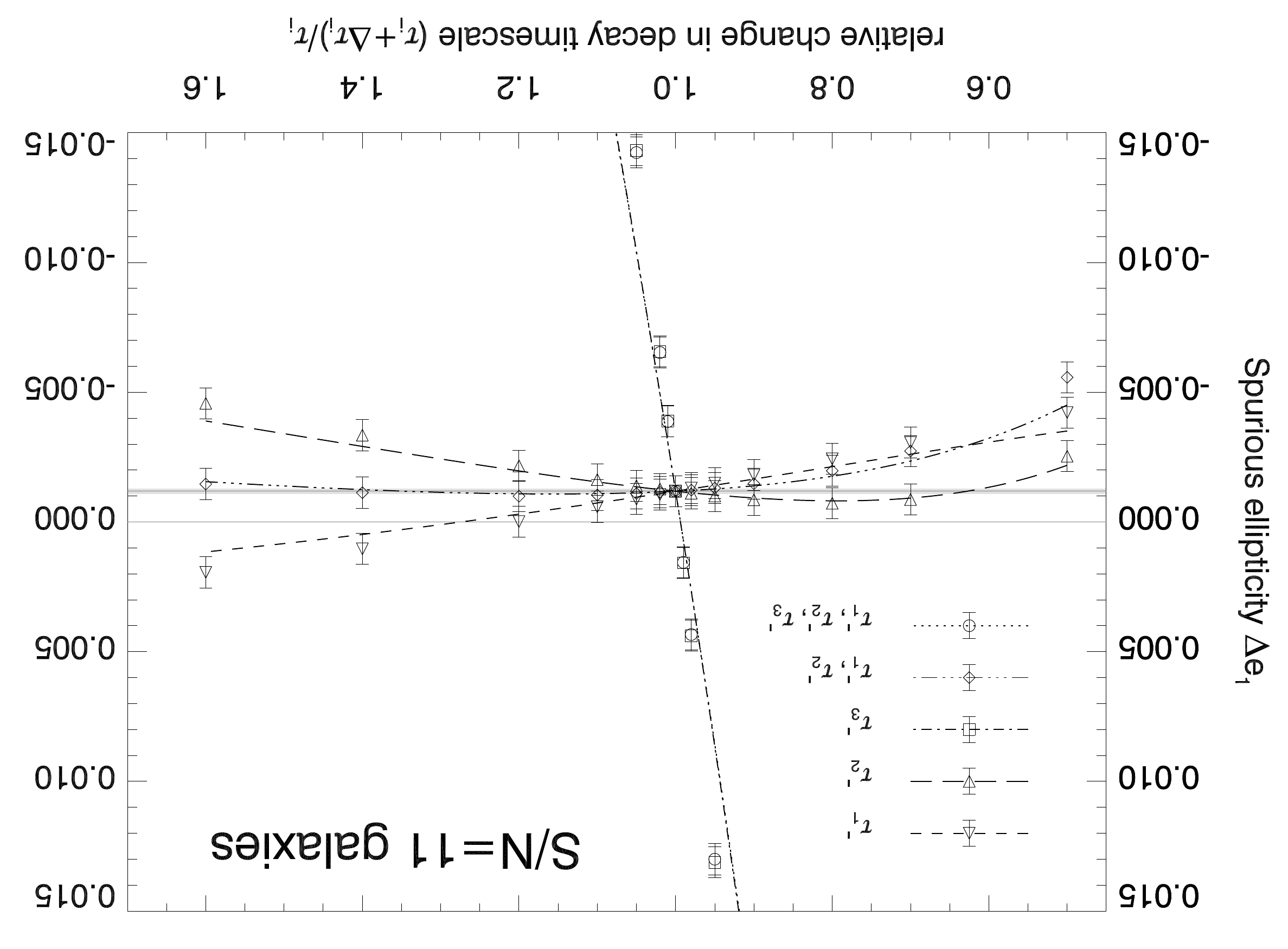}
\includegraphics[width=140mm,angle=180]{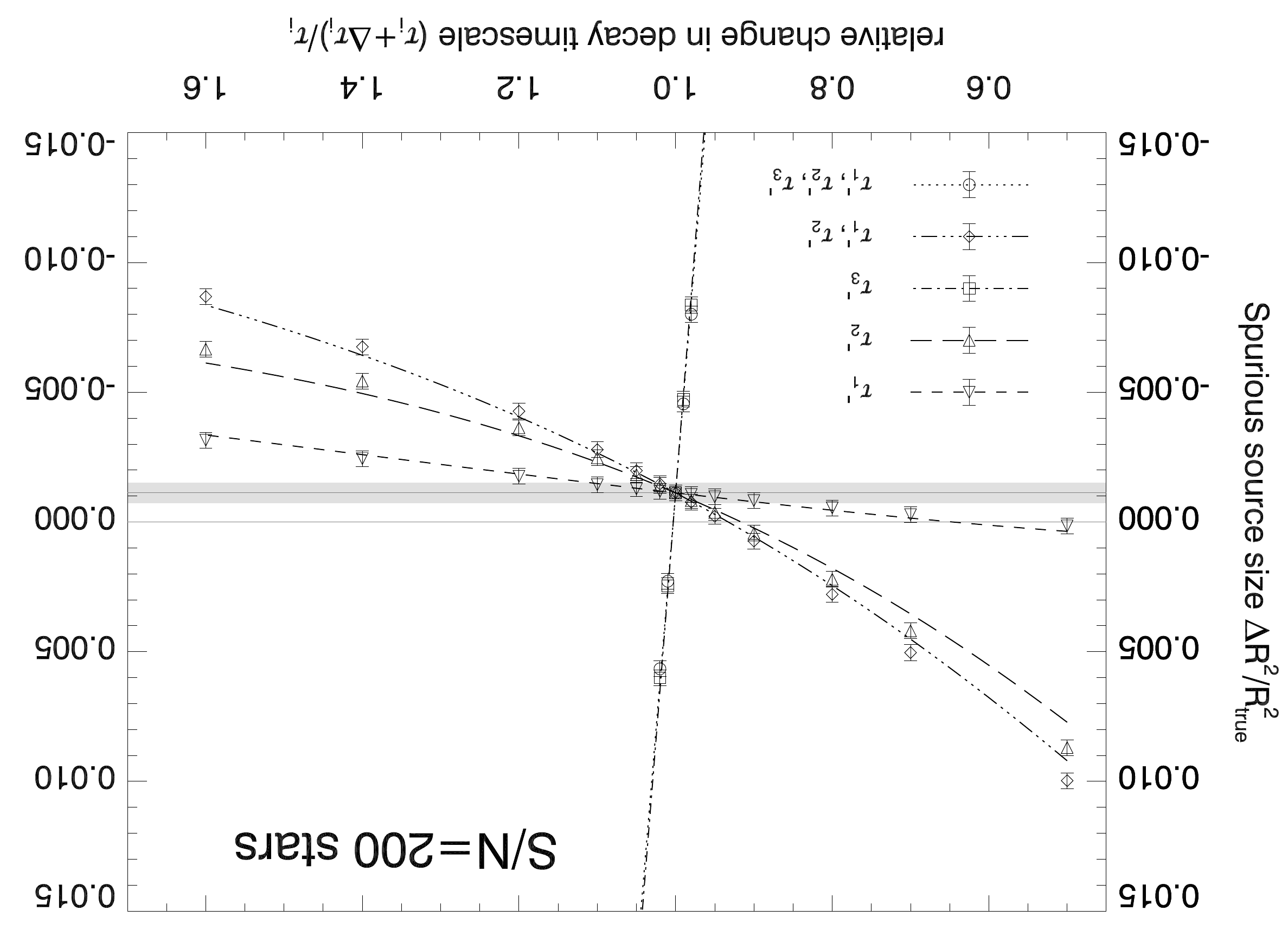}
\caption{Sensitivity of the CTI-induced spurious ellipticity $\Delta e_{1}$ 
in faint galaxies 
\textit{(upper panel)} and relative bias in source size 
$\Delta R^{2}/R^{2}_{\mathrm{true}}$ \textit{(lower panel)} 
in bright stars to a relative change in release timescales 
$(\tau_{i}+\Delta\tau_{i})/\tau_{i}$. Different 
symbols and line styles denote to which of the trap species a change in 
release timescale was applied: The slow traps: ($\tau_{1}$, upward 
triangles and dashed line); the medium traps: ($\tau_{2}$, downward
triangles and long-dashed line); both of them: ($\tau_{1}$, $\tau_{2}$,
diamonds and triple dot-dashed line); the fast traps ($\tau_{3}$, 
squares and dot-dashed line); all: ($\tau_{1}$, $\tau_{2}$, $\tau_{3}$, 
circles and dotted line). The various broken lines 
show the best-fit representation of the
measurements as given by the empiric sensitivity model 
(eq.~\ref{eq:taupred}). Like in Fig.~\ref{fig:mdm2_exp0}, the grey shaded area
indicates the VIS \textit{Euclid} requirement range. 
We study the worst affected objects (at the end of the mission
and furthest from the readout register) and the faintest \textit{Euclid}
galaxies.}
\label{fig:mdm2_exp2}
\end{figure*}
\begin{table*}
 \centering
 \caption{Tolerances for changes in the trap model parameters $\xi$ (column 1),
 derived from polynomial fits to the sensitivity curves, and taking into account the
 \textit{Euclid} VIS requirements (eqs.~\ref{eq:polynom}, \ref{eq:tol}). 
 Here, we assume no residual biases when using the correct trap model. 
 For three observables $\eta$, we show pairs of tolerances, for $\Delta\xi\!<\!0$ and 
 $\Delta\xi\!>\!0$ each. Columns 2 and 3 report
 tolerances based on the ellipticity bias $\Delta e_{1}$, columns 4 and 5 those from the relative
 size bias $\Delta R^{2}/R^{2}_{\mathrm{true}}$, and columns 6 and 7 those from the relative
 flux bias $\Delta F/F_{\mathrm{true}}$.}
 \begin{tabular}{cccccccc}
 \hline\hline
  branch $\xi$ & $10^{4}\Delta\xi_{\mathrm{tol}}^{\mathrm{min}}$ & 
  $10^{4}\Delta\xi_{\mathrm{tol}}^{\mathrm{max}}$ &
  $10^{4}\Delta\xi_{\mathrm{tol}}^{\mathrm{min}}$ &
  $10^{4}\Delta\xi_{\mathrm{tol}}^{\mathrm{max}}$ &
  $10^{4}\Delta\xi_{\mathrm{tol}}^{\mathrm{min}}$ &
  $10^{4}\Delta\xi_{\mathrm{tol}}^{\mathrm{max}}$ \\
   & $\eta\!=\!\Delta e_{1}$ & $\eta\!=\!\Delta e_{1}$ & 
  $\eta\!=\!\Delta R^{2}/R^{2}_{\mathrm{true}}$ & $\eta\!=\!\Delta R^{2}/R^{2}_{\mathrm{true}}$ & 
  $\eta\!=\!\Delta F/F_{\mathrm{true}}$ & $\eta\!=\!\Delta F/F_{\mathrm{true}}$ \\
 & galaxies & galaxies & stars & stars & galaxies & galaxies \\ \hline
  $\beta$ & $-0.631\pm0.007$ & $0.631\pm0.007$ & $-4.78\pm0.05$ & $4.78\pm0.05$ & $-61.5\pm0.3$ & $60.5\pm0.3$ \\
  $\rho_{1}$ & $-84_{-33}^{+18}$ & $84_{-18}^{+33}$ & $-1250_{-1800}^{+450}$ & $1250_{-450}^{+1800}$ & $--$ & $--$ \\
  $\rho_{2}$ & $-39_{-6}^{+4}$ & $39_{-4}^{+6}$ & $-191_{-19}^{+16}$ & $191_{-16}^{+19}$ & $--$ & $--$ \\
  $\rho_{3}$ & $-3.03_{-0.07}^{+0.06}$ & $3.03_{-0.06}^{+0.07}$ & $-5.91\pm0.03$ & $5.91\pm0.03$ & $-267.5\pm1.6$ & $267.5\pm1.6$ \\
  $\rho_{1,2}$ & $-26_{-3}^{+2}$ & $26_{-2}^{+3}$ & $-166_{-14}^{+12}$ & $166_{-12}^{+14}$ & $--$ & $--$ \\
  $\rho_{1,2,3}$ & $-2.72\pm0.05$ & $2.72\pm0.05$ & $-5.71\pm0.03$ & $5.71\pm0.03$ & $-262.8\pm1.6$ & $262.8\pm1.6$ \\
  $\tau_{1}$ & $-193_{-23}^{+19}$ & $193_{-19}^{+23}$ & $-1310_{-150}^{+120}$ & $1310_{-120}^{+150}$ & $<-10000$ & $>10000$ \\
  $\tau_{2}$ & $-300_{-360}^{+90}$ & $270_{-70}^{+150}$ & $-270_{-70}^{+50}$ & $270_{-50}^{+80}$ & $<-10000$ & $>10000$ \\
  $\tau_{3}$ & $-4.00\pm0.04$ & $4.00\pm0.04$ & $-11.30\pm0.05$ & $11.31\pm0.05$ & $-1574_{-23}^{+24}$ & $2320_{-90}^{+100}$ \\
  $\tau_{1,2}$ & $-420_{-420}^{+150}$ & $700_{-400}^{+900}$ & $-220_{-50}^{+30}$ & $230_{-40}^{+50}$ & $<-10000$ & $>10000$ \\
  $\tau_{1,2,3}$ & $-4.03\pm0.04$ & $4.04\pm0.04$ & $-11.69\pm0.05$ & $11.68\pm0.05$ & $-1454_{-20}^{+19}$ & $2020_{-60}^{+70}$ \\
$\tau_{1,2,3}, \rho_{1,2,3}$, first pixel matched & $-16.07_{-0.61}^{+0.57}$ & $16.09_{-0.57}^{+0.61}$ & $-16.17\pm0.09$ & $16.21\pm0.09$ & $-262.5\pm0.7$ & $263.0\pm0.7$ \\
  \hline\hline\label{tab:tolerances}
  \vspace{-5.5mm}
 \end{tabular}
\end{table*}
Figure~\ref{fig:mdm2_exp2} shows the $\mathbf{M}+\delta\mathbf{M}$ experiment
for one or more of the release timescales $\tau_{i}$ of the trap model.
The upper panel of Fig.~\ref{fig:mdm2_exp2} presents the spurious ellipticity
$\Delta e_{1}$ for five different branches of the experiment.
In each of the branches, we modify the release timescales $\tau_{i}$ of one or 
several of the trap species by multiplying it with a factor $(\tau_{i}+\Delta\tau_{i})/\tau_{i}$. 

As in Fig.~\ref{fig:mdm2_exp1}, the upward triangles in 
Fig.~\ref{fig:mdm2_exp2} denote that the correction model applied to the 
images degraded with the baseline model used a density of 
$\tau_{1}+\Delta\tau_{1}$ 
for the fast trap species. The release timescales of the other species are 
kept to their baseline values in this case. The other four branches modify 
$\tau_{2}$ (downward triangles); $\tau_{3}$ (squares); $\tau_{1}$ 
and $\tau_{2}$ (diamonds); and all three trap species (circles).

Because a value of $\Delta\tau\!=\!0$
reproduces the baseline model in all branches,
all of them recover the zeropoint measurement of $\Delta e_{1}$ there.
The three trap species differ in how the $\Delta e_{1}$ they induce varies as 
a function of $\Delta\tau_{i}$.
One the one hand, for $\tau_{1}$, we observe more 
negative $\Delta e_{1}$ for $(\tau_{i}+\Delta\tau_{i})/\tau_{i}\!<\!1$ , 
and less negative values for 
$(\tau_{i}+\Delta\tau_{i})/\tau_{i}\!>\!1$, with a null at 
$(\tau_{i}+\Delta\tau_{i})/\tau_{i}\!\approx\!1.5$. On the other hand,
with the slow traps ($\tau_{3}$), we find $\Delta e_{1}\!>\!0$ for 
$(\tau_{i}+\Delta\tau_{i})/\tau_{i}\!\la\!0.99$, and more negative values than the zeropoint for 
$(\tau_{i}+\Delta\tau_{i})/\tau_{i}\!>\!1$. The curve of $\Delta e_{1}(\lambda\tau_{2})$ shows a maximum at 
$(\tau_{i}+\Delta\tau_{i})/\tau_{i}\!\approx\!0.8$, with a weak dependence on 
$0.7\!\la\!(\tau_{i}+\Delta\tau_{i})/\tau_{i}\!\la\!1.1$. 

Key to understanding the spurious ellipticity as a function of the $\tau_{i}$
is the dependence of $\Delta e_{1}(\tau)$ for a single trap species that we
presented in Fig.~\ref{fig:raw4panel}, and expressed by the empirical fitting
function $f_{\mathrm{e_{\alpha}}}(\tau)$ (Eq.~\ref{eq:adg}) with the parameters 
quoted in Table~\ref{tab:tolerances}.
While the correction algorithm effectively removes the trailing when the true
$\tau_{i}$ is used, the residual of the correction will depend on the 
difference between the $\Delta e_{\alpha}$ for $\tau_{i}$ and for the timescale 
$(\tau_{i}+\Delta\tau_{i})/\tau_{i}$ actually used in the correction. 
This dependence is captured by the predictive model (Eq.~\ref{eq:sumpred}), which
simplifies for the situation in Fig.~\ref{fig:mdm2_exp2} ($\Delta\rho_{i}\!=\!0$) to
\begin{equation} \label{eq:taupred}
\Delta f^{\mathrm{Pr}}(\tau_{i}+\Delta\tau_{i})\!=\!Z+
\sum_{i}\rho_{i}\left[f(\tau_{i}) - f(\tau_{i}+\Delta\tau_{i})\right],
\end{equation}
with $Z\!=\!\sum_{i}\rho_{i}f^{\mathrm{resid}}(\tau_{i})$
(lines in Fig.~\ref{fig:mdm2_exp2}). 
In the branches modifying $\tau_{1}$ and/or $\tau_{2}$, but not $\tau_{3}$,
the measurements over the whole range of 
$0.5\!\leq\!(\tau_{i}+\Delta\tau_{i})/\tau_{i}\!\leq\!1.6$ 
agree with the empirical model within their uncertainties. 
If $\tau_{3}$ is varied, Eq.~(\ref{eq:taupred}) overestimates $|\Delta e_{1}|$ significantly for 
$|\Delta\tau_{i}|\!>\!0.05\tau_{i}$. We
discuss a possible explanation in Sect.~\ref{sec:disc}. 
Our empirical model provides a natural explanation for the maximum in 
$\Delta e_{1}(\tau_{2})$: Because $\tau_{2}\!=\!3.5$ is located near the peak in
$f_{\mathrm{e_{1}}}(\tau)$, assuming 
$(\tau_{i}+\Delta\tau_{i})/\tau_{i}\!\leq\!0.8$ for correction means
using a release time regime where $\Delta e_{1}(\tau)$ is still rising instead of
falling. The correction software accounts for this; hence the spurious ellipticity
from using the wrong release time scale shows the same maximum as 
$f_{\mathrm{e_{1}}}(\tau)$.

Because $\tau_{2}$ is not located very closely to the peak in 
$\Delta R^{2}/\Delta R^{2}_{\mathrm{true}}(\tau)$ (cf.\ Fig.~\ref{fig:raw4panel}),
we do not see an extremum in the lower panel of Fig.~\ref{fig:mdm2_exp2}
which shows the sensitivity of the size bias in bright stars to variations 
in the $\tau_{i}$. 

In order to compute the tolerances $\Delta\tau_{\mathrm{tol}}$ 
towards changes in the release timescales, we again employ a polynomial fit
(eq.~\ref{eq:tol}). 
Evidently, the tolerances differ substantially between the $\tau_{i}$, again with the
narrower tolerance intervals from $\Delta e_{1}$ than from 
$\Delta R^{2}/\Delta R^{2}_{\mathrm{true}}$. 
Only for $\Delta\tau_{2}$ with its extreme point for $\Delta e_{1}$ 
near the baseline value, we find similar tolerances in both cases. 
However, even for the rare trap species $\tau_{1}$, the tolerance is only 
$\Delta\tau_{1,\mathrm{tol}}\!=\!(1.93\pm0.23)$~\%. 
One needs to know the release timescale of
the slow trap species to an accuracy of 
$(0.0400\pm0.0004)$~\% 
to be able to correct it within \textit{Euclid} VIS requirements. 
We find the same tolerance if all timescales are varied in unison.

\subsubsection{Combinations of timescales and densities yielding the same first trail pixel flux} \label{sec:sti}

\begin{figure}
\includegraphics[width=90mm,angle=180]{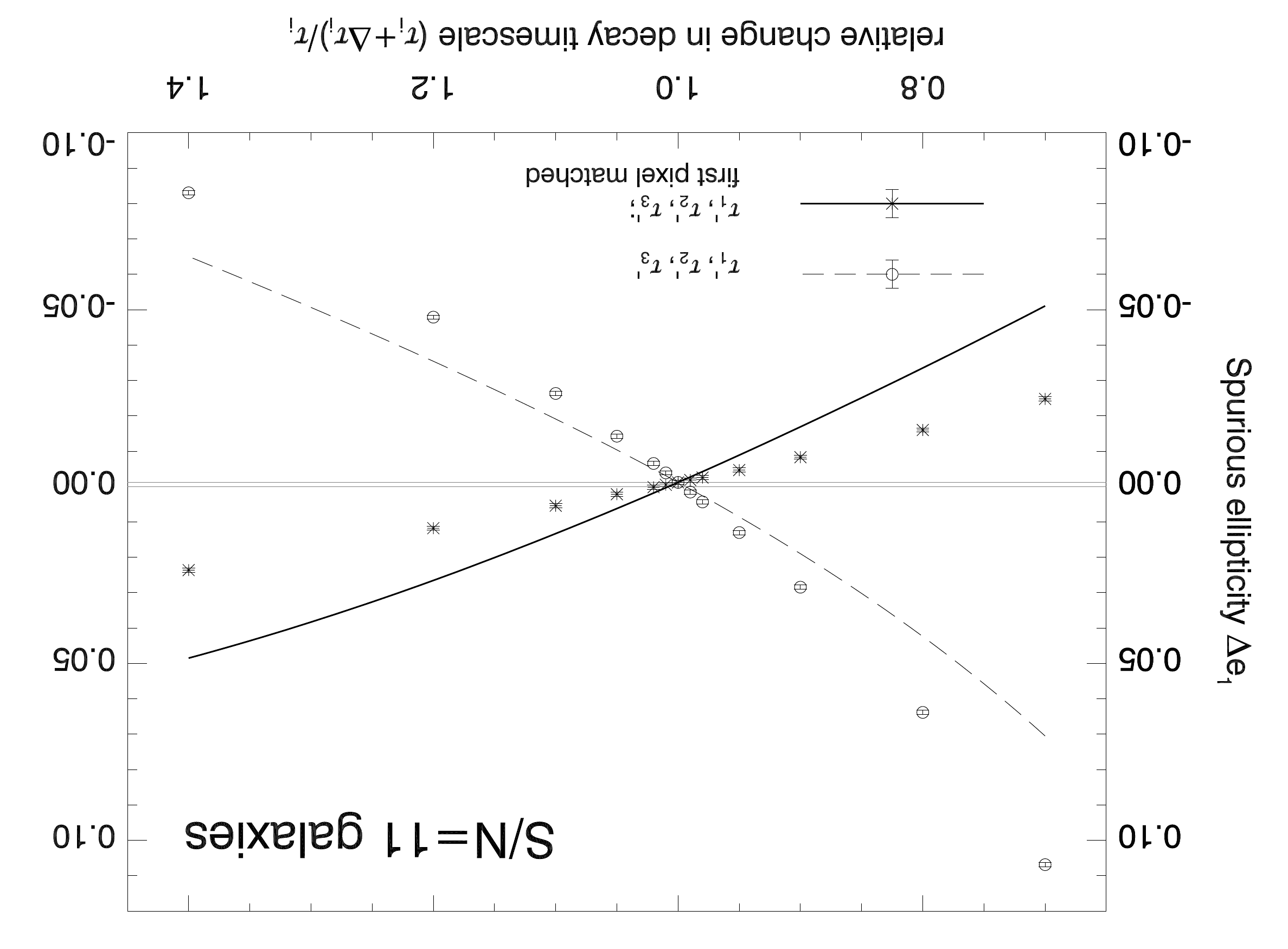}
\includegraphics[width=90mm,angle=180]{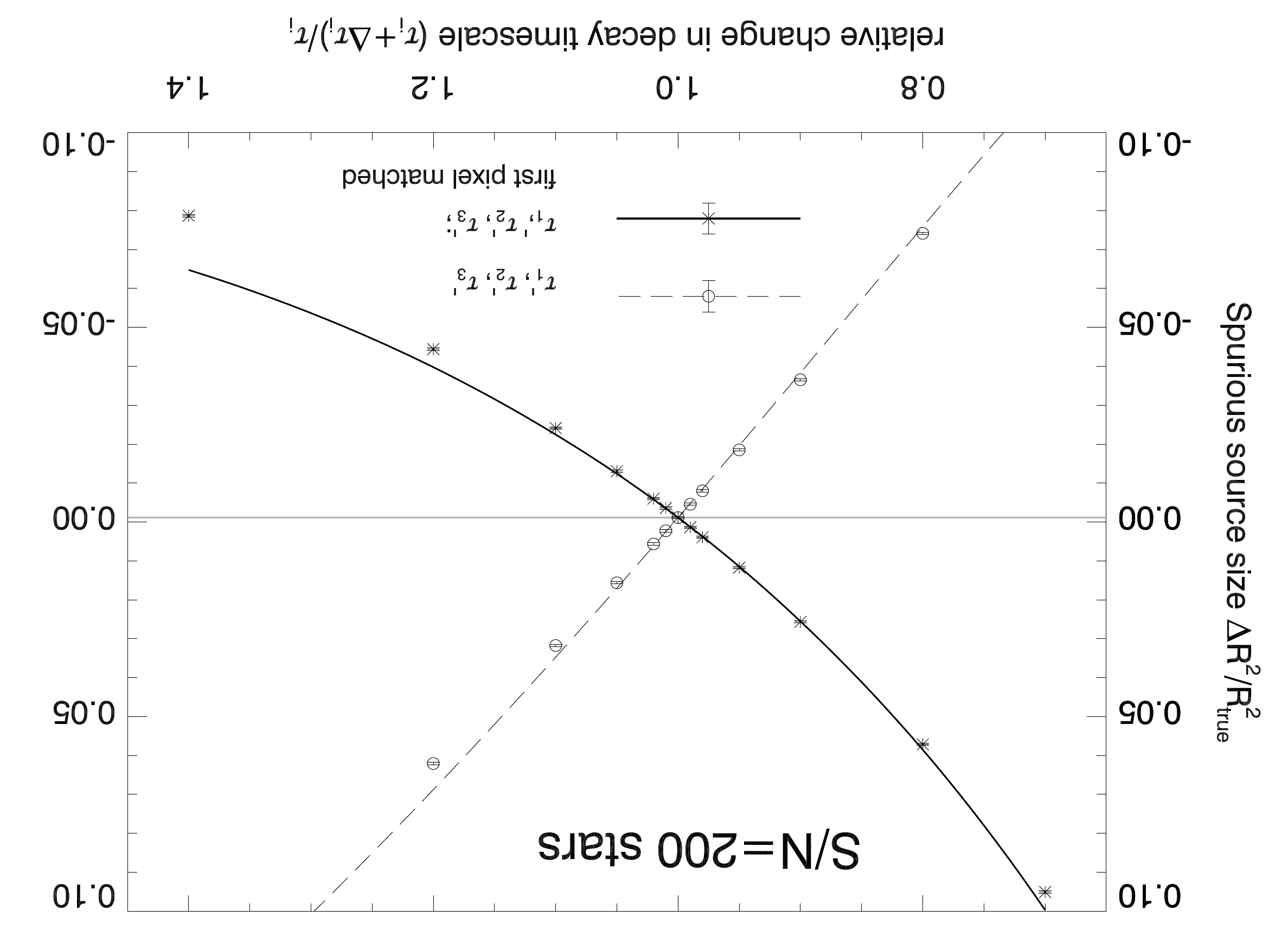}
\caption{The same as Fig.~\ref{fig:mdm2_exp2}, but for $\Delta\tau_{i}\!<\!0$ 
combinations of timescales $\tau_{i}$ and densities $\rho_{i}$ 
that yield the same count rate in the first trail pixel as the baseline model.
All trap species are modified in unison (large symbols and solid line).
For comparison, small symbols and the dotted line repeat the result from 
Fig.~\ref{fig:mdm2_exp2}, where only the $\tau_{i}$ were modified, 
not the $\rho_{i}$. (Notice the different scale of the ordinates.) 
The lines show the predictive models (Eq.~\ref{eq:sumpred}). 
We study the worst affected objects (end of the mission,
furthest from the readout register) and the faintest \textit{Euclid}
galaxies.}
\label{fig:mdm2_exp3}
\end{figure}
Considering how trap parameters are constrained practically from 
Extended Pixel Edge Response (EPER) and First Pixel Response (FPR) 
data, it is instructive to consider combinations of trap release timescales 
$\tau_{i}$ and densities $\rho_{i}$ that yield the same number of electrons 
in the first pixel of the trail as the baseline model.
This is interesting because given realistic conditions, the first pixel of the
trail will have the largest signal-to-noise ratio and will be most easily constrained. 
We thus perform an initial exploration of the parameter degeneracies.
In our ``first pixel matched'' models, the effect of
a given change in $\tau$ on the first trail pixel needs to be compensated
by a change in $\rho$.
Because a larger (smaller) $\tau$ means more (less) 
charge close to the original pixel, the compensation requires 
$\Delta\rho_{i}\!<\!0$ for $\Delta\tau_{i}\!<\!1$ and
$\Delta\rho_{i}\!>\!1$ for $\Delta\tau_{i}\!>\!1$. 
Only in the branches where we vary $\tau_{3}$ or all timescales together,
we find the $\Delta\rho_{i}$ to differ 
noticeably from unity. For the latter two, they populate a range between 
$\Delta\rho_{i}\!=\!0.745$ for $\Delta\tau_{i}\!=\!0.7$ to 
$\Delta\rho_{i}\!=\!1.333$ for $\Delta\tau\!=\!1.4$. 

Figure~\ref{fig:mdm2_exp3} shows the $\mathbf{M}+\delta\mathbf{M}$ experiment
for all $\tau_{i}$ and $\rho_{i}$ (large symbols). Small symbols depict the
alteration to $\tau_{i}$, but with the $\rho_{i}$ kept fixed, i.e.\ the same
measurement as the open circles in Fig.~\ref{fig:mdm2_exp2}. Compared to these,
$\Delta e_{1}$ 
in faint galaxies 
(upper panel)
is of opposite sign in the ``first pixel matched'' case, relative to the zeropoint.
This can be understood as an effect of our baseline trap mix being dominated 
by slow traps, for which a small increase in $\tau$ leads to \emph{less}
CTI-induced ellipticity. The simultaneous increase in trap density effects
\emph{more} CTI-induced ellipticity, and this is the larger of the two terms,
such that a change in sign ensues. 
The same holds for $\Delta R^{2}/R^{2}_{\mathrm{true}}$ 
in bright stars 
(lower panel of Fig.~\ref{fig:mdm2_exp3}), 
but with inverted slopes compared to $\Delta e_{1}$. 

Again using eq.~(\ref{eq:tol}), we compute the tolerance range for the changes to the
$\tau_{i}$ in the ``first pixel matched'' case. (The respective changes to the
$\rho_{i}$ are determined by the first pixel constraint.) 
Modifying all release time scales, we arrive at 
$\Delta\tau_{\mathrm{tol}}\!=\!0.16$~\%. 
(Table~\ref{tab:tolerances}). 
This tolerance is wider than the 
$0.04$\% for $\Delta e_{1}$ 
when only the $\tau_{i}$ are varied, again due to the different signs 
arising from variations to $\tau_{3}$ and $\rho_{3}$. 
By coincidence, we also arrive at $\Delta\tau_{\mathrm{tol}}\!=\!0.16$~\%
when repeating that test with the size bias measured in bright stars. 

The black solid line in Fig.~\ref{fig:mdm2_exp3} shows the predictive
model (eq.~\ref{eq:sumpred}), taking into account the combined effect of the
$\Delta\tau_{i}$ and $\Delta\rho_{i}$, giving the same first pixel flux. 
Both in the $\tau_{i}$-only (dotted line) and ``first pixel matched'' cases 
it matches the measurements only 
within a few percent from $\lambda\!=\!1$. 
Crucially, this mismatch only occurs for $\Delta e_{1}$ in faint galaxies, 
but not for $\Delta R^{2}/R^{2}_{\mathrm{true}}$ in bright stars. 

We explain this discrepancy with the uncertainties with which our
measurements and modelling (Fig.~\ref{fig:raw4panel}) describe the
underlying function $f_{\mathrm{e_{1}}}(\tau)$.
The range $20\!\la\!\tau\!\la\!100$
is where the fitting function Eq.~(\ref{eq:adg}) deviates most from the
observations in Fig.~\ref{fig:raw4panel}. The CTI correction effectively 
removes almost all CTI effects on photometry and morphology, leaving the
residuals presented in Figs.~\ref{fig:mdm2_exp1} to \ref{fig:flux2}, at least
one order of magnitude smaller than the scales of the uncorrected CTI effects.
Hence, a relatively small uncertainty in 
$f(\tau)$ causes a large mismatch with the data.

The cause of the uncertainty in the parameters of Eq.~(\ref{eq:adg}), shown
in Table~\ref{tab:taufits}, is twofold: First, there is uncertainty in the fit
as such. Second, there is uncertainty due to the finite sampling of 
the $\Delta e_{\alpha}(\tau)$ and $\Delta F_{\mathrm{rel}}(\tau)$ curves.
Running a denser grid in $\tau$ can remove the latter, but the former might
be ultimately limited by our choice of the function (Eq.~\ref{eq:adg}), 
which is empirically motivated, not physically.
We further discuss the limits of the predictive model in Sect.~\ref{sec:disc}.

\subsection{Residual flux errors after imperfect CTI correction} \label{sec:photo}

\subsubsection{Flux bias as a function of readout noise} \label{sec:fluxrn}

\begin{figure}
\includegraphics[width=90mm,angle=180]{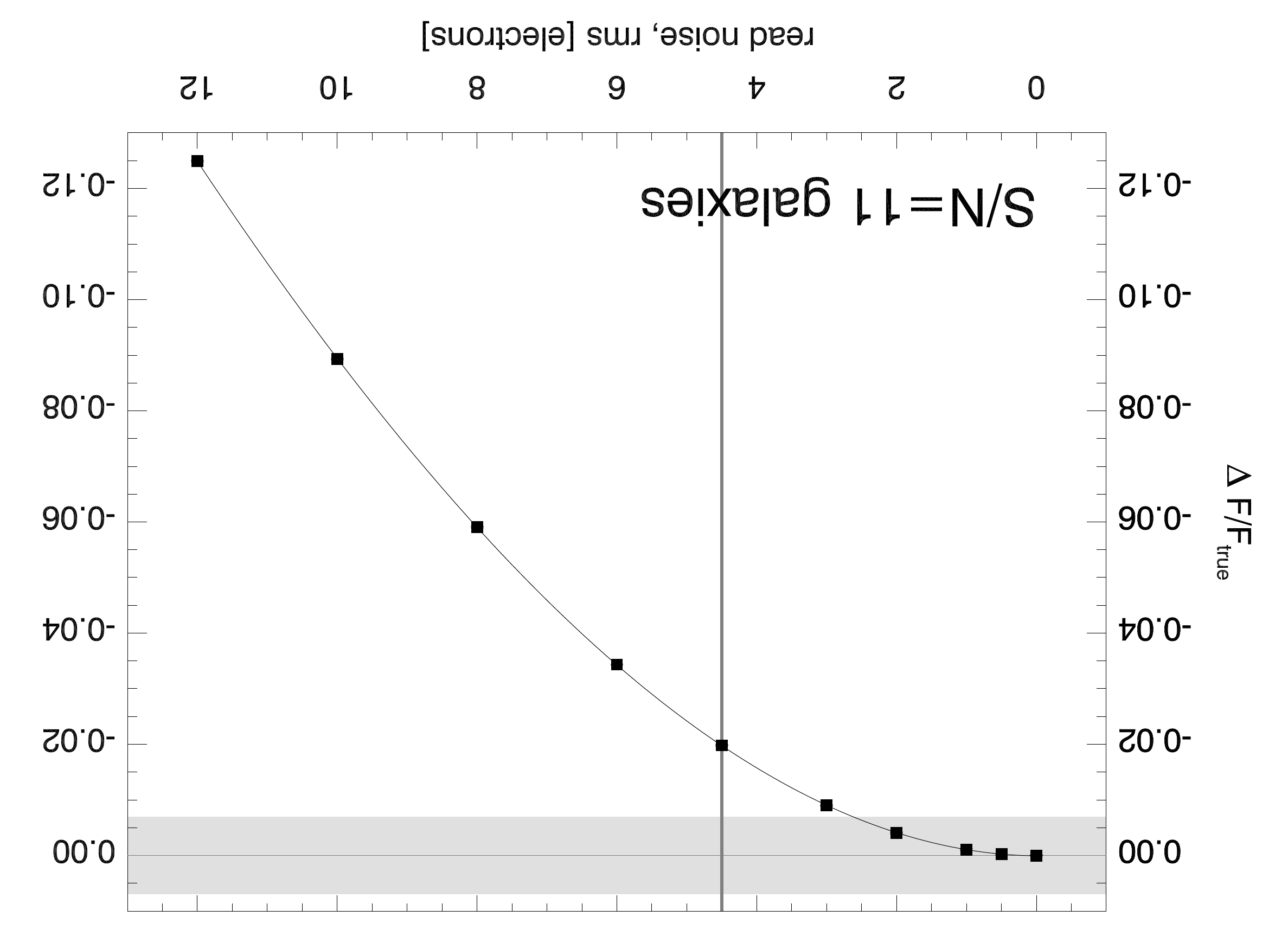}
\includegraphics[width=90mm,angle=180]{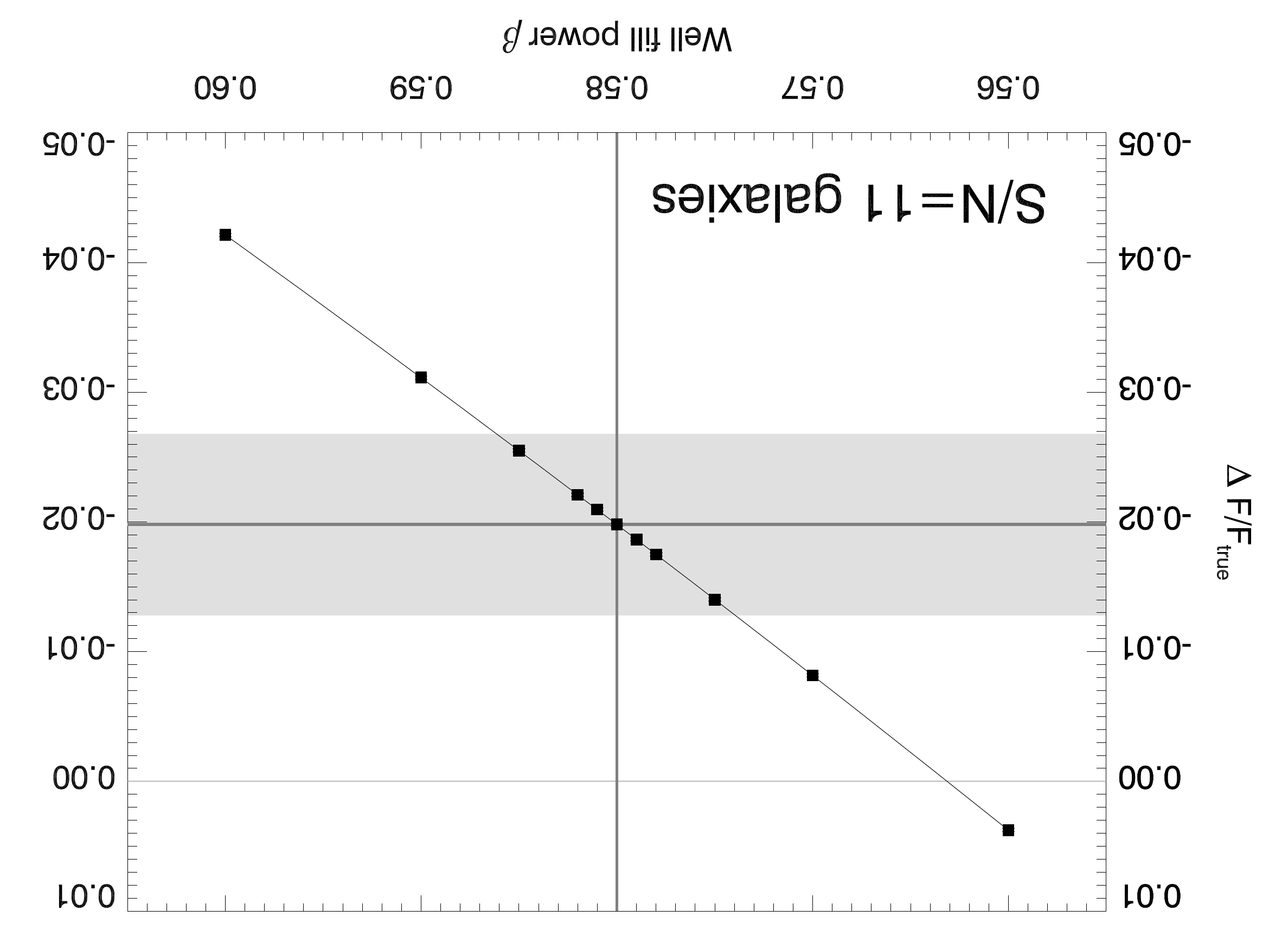}
\caption{Relative bias in RRG flux with respect to the true input flux,
as a function of readout noise (\textit{upper panel}) and well fill power $\beta$
(\textit{lower panel}). Solid lines give the best-fit polynomial models. 
The grey-shaded \textit{Euclid} requirement
range is centred on zero for the readout noise plot, and on the zeropoint
corresponding to the default readout noise for the $\beta$ plot.
Measurement uncertainties are shown, but very small. 
We study the worst affected objects (end of the mission,
furthest from the readout register) and the faintest \textit{Euclid}
galaxies.}
\label{fig:flux1}
\end{figure}
Given the default rms readout noise of $4.5$ electrons, we measure a flux bias
$\Delta F_{\mathrm{rel}}\!=\!\Delta F/F_{\mathrm{true}}$ relative to the true flux 
$F_{\mathrm{true}}$ in the input
faint galaxy simulations of $(-1.980\pm0.012)$~\% 
after CTI correction, corresponding to $92.9$~\% of the CTI-induced flux bias
being corrected.
The upper panel of Fig.~\ref{fig:flux1} shows the relative 
flux biases before and after correction
as a function of rms readout noise. Without readout noise, 
the flux bias can be corrected perfectly
($\Delta F_{\mathrm{rel}}\!=\!(0.002\pm0.012)\times 10^{-2}$ 
after correction). With increasing readout noise, the flux bias 
deteriorates, in a way that can be fitted with a cubic polynomial in 
terms of readout noise. 
Comparing to the degraded images, we notice that the correction software 
applies same amount of correction, 
independent of the readout noise. Because the mitigation algorithm in its current form does not
include a readout noise model, this confirms our expectations.

We show the \textit{Euclid} requirement on photometric accuracy as the 
grey-shaded area in Fig.~\ref{fig:flux1} (upper panel), centred on zero.
The nominal readout noise case exceeds 
the requirement of $<\!0.7$~\% photometric uncertainty 
for the faintest, worst-affected galaxies we study.  
However, the CTI-induced bias affects all VIS images, and would thus be
calibrated out. The \textit{Euclid} flux requirement can be understood as
pertaining to \textit{uncertainties}, not \textit{biases} in the
photometric calibration. The uncertainty of the flux bias, 
$0.0012$~\% 
then makes only a tiny contribution to the photometric error budget. 
We now go on to study the sensitivity of the flux bias towards changes in the trap model.

\subsubsection{Flux bias as a function of well fill power $\beta$} 
 
The lower panel of Fig.~\ref{fig:flux1} shows how a change in well fill
power $\beta$ alters the flux bias. If we correct the degraded images using a
$\beta\!>\!\beta_{0}$, the model accounts for less CTI in small 
charge packages, i.e.\ 
less CTI in the image's wings that are crucial for both
photometry and morphology (cf.\ fig.~\ref{fig:mdm2_exp0}) 
Hence, a $\beta\!>\!\beta_{0}$ leads to an 
undercorrection relative to the flux bias zeropoint $Z_{\mathrm{F}}$
(Sect.~\ref{sec:fluxrn}), while for $\beta\!-\!\beta_{0}\!\la\!-0.017$, 
the zero line is crossed and overcorrection occurs. 

Although $\Delta F_{\mathrm{rel}}(\beta)$ in Fig.~\ref{fig:flux1} appears 
linear, using the criterion based on significant components (Sect.~\ref{sec:rho}),
a quadratic is preferred, 
indicated by the solid line. Using eq.~(\ref{eq:tol}), 
we compute the tolerance range in
$\beta$ given $\Delta F_{\mathrm{rel}}(\beta_{\mathrm{tol}})\!=\!0.007$,
centred on 
$Z_{\mathrm{F}}$. Towards smaller well fill powers, we find 
$\Delta\beta_{\mathrm{tol}}^{\mathrm{min}}\!=\!-(6.15\pm0.03)\times 10^{-3}$, 
while towards larger $\beta$, we find 
$\Delta\beta_{\mathrm{tol}}^{\mathrm{max}}\!=\!(6.05\pm0.03)\times 10^{-3}$. 
Compared to the constraints on the knowledge of $\beta$ from $\Delta e_{1}$
derived in Sect.~\ref{sec:beta}, these margins are 
$\sim\!100$ times wider.

\subsubsection{Flux bias as a function of trap densities} \label{sec:fluxrho}

The upper plot of Fig.~\ref{fig:flux2} shows the flux bias 
$\Delta F_{\mathrm{rel}}$ in dependence of a change $\Delta\rho_{i}$ 
to the densities $\rho_{i}$ in the correction model, in analogy to Sect.~\ref{sec:rho}.
Unless the density of the dominant trap species $\rho_{3}$ is modified, 
we measure only 
insignificant departures 
from the zeropoint $Z_{\mathrm{F}}$. Given the high accuracy of the flux measurements,
these are still significant measurements, but they are negligible with respect 
to the \textit{Euclid} requirement on flux. If all $\rho_{i}$ are varied in
unison, the effect on $\Delta F_{\mathrm{rel}}$ is largest.
A linear model using Eq.~(\ref{eq:polynom}) yields a tolerance of 
$\Delta\rho_{i}^{\mathrm{tol}}/\rho_{i}\!=\!\pm2.628\pm0.016$~\%,
wider than the tolerances for $\Delta e_{1}$ 
or $\Delta R^{2}/R^{2}_{\mathrm{true}}$ (Table~\ref{tab:tolerances}).
The lines in the upper plot of Fig.~\ref{fig:flux2} show that the
model (eq.~\ref{eq:rhopred}) matches our measurements well over the range in
$\Delta\rho_{i}$ we tested.

\subsubsection{Flux bias as a function of release timescales}

\begin{figure}
\begin{center}
\vspace{-0.54cm}
\includegraphics[width=90mm,angle=180]{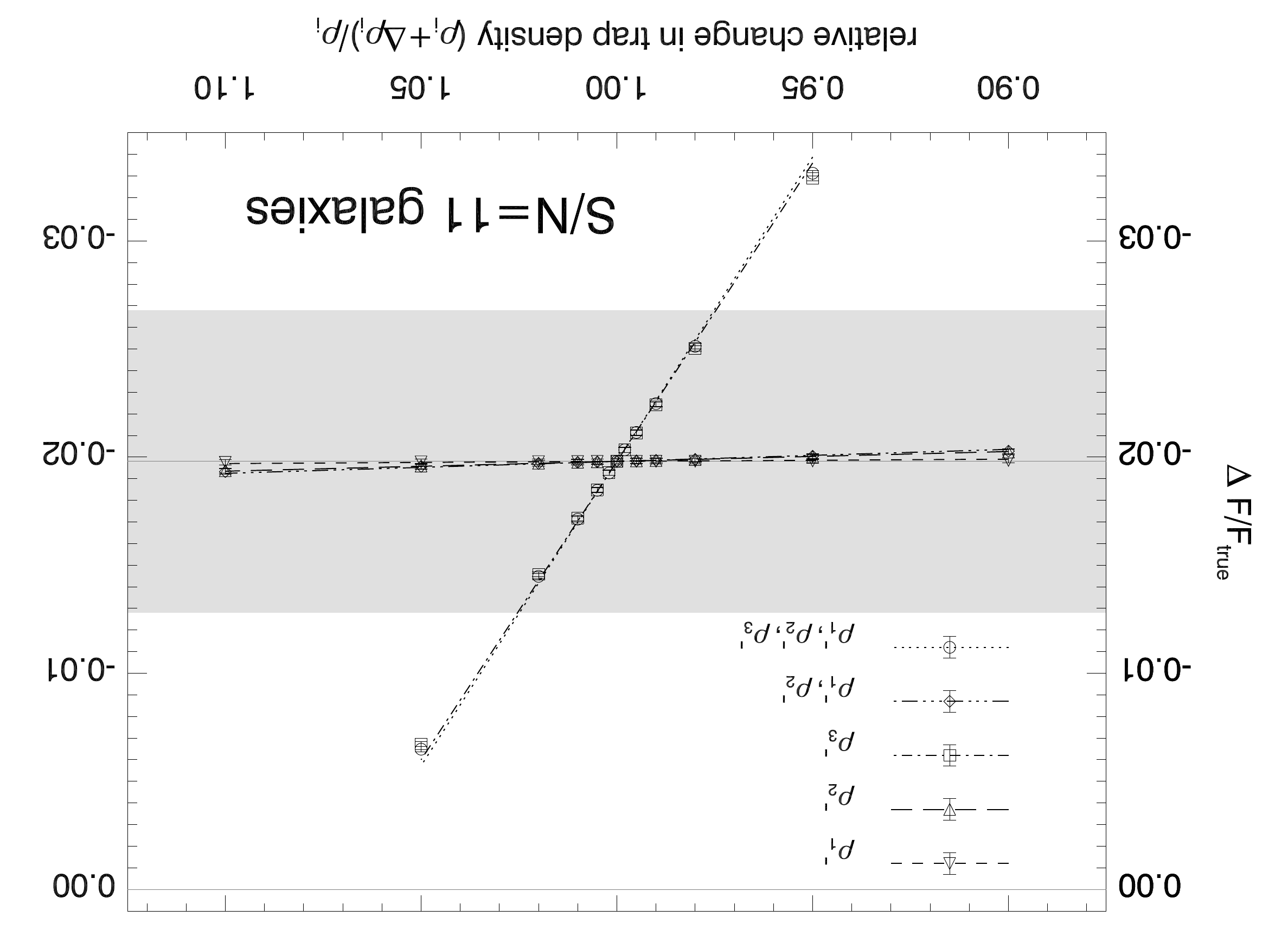}
\includegraphics[width=90mm,angle=180]{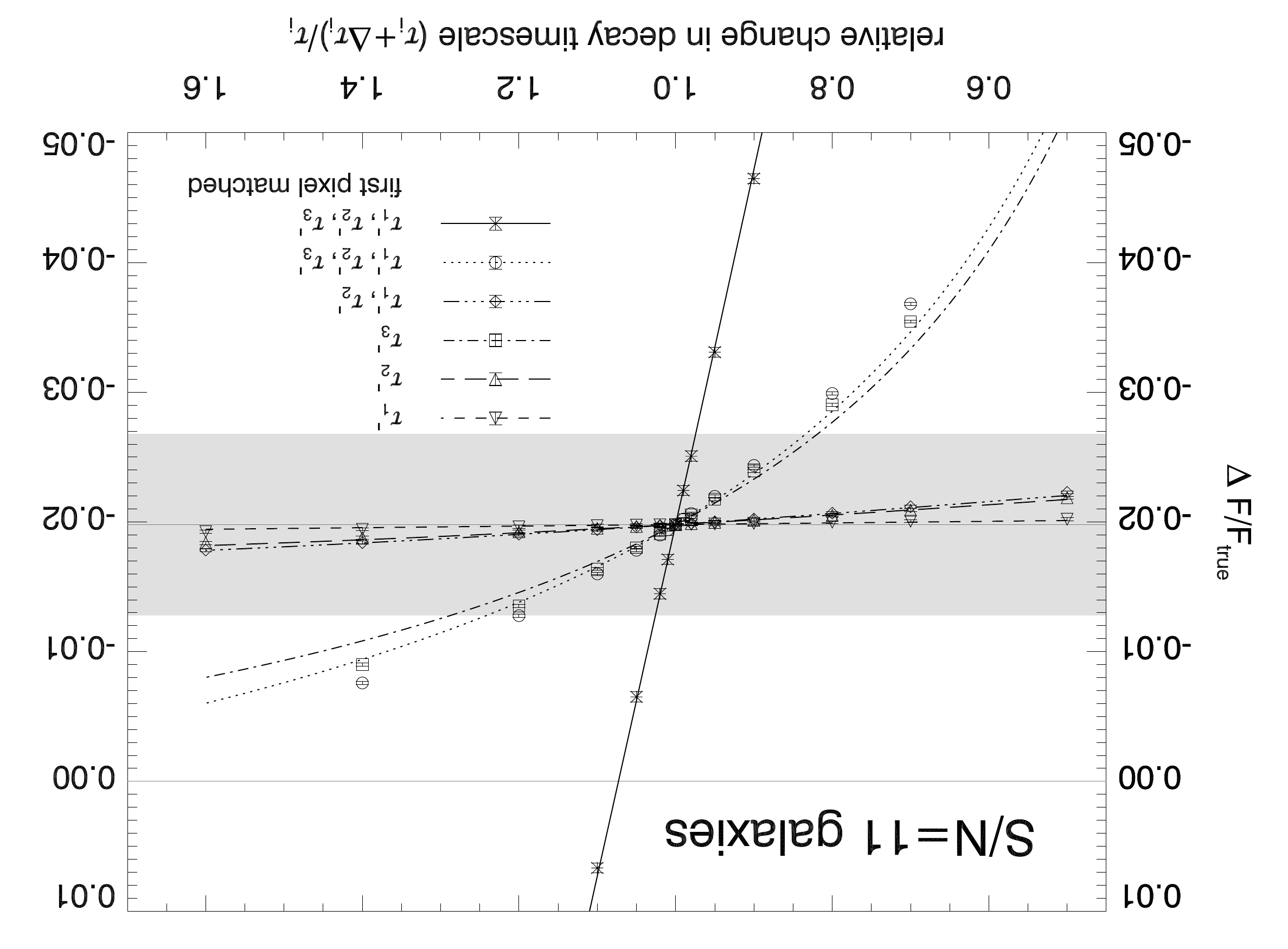}
\end{center}
\caption{\textit{Upper panel:} The same as 
Fig.~\ref{fig:mdm2_exp1}, but showing the sensitivity of the measured flux bias
$\Delta F/F_{\mathrm{true}}$ as a function of the relative change in 
trap densities $\rho_{i}$. 
\textit{Lower panel:} The same as 
Fig.~\ref{fig:mdm2_exp2}, but showing the flux bias
$\Delta F/F_{\mathrm{true}}$ as a function of the relative change in 
trap densities $\tau_{i}$. Star symbols and the solid line
denote the ``first pixel matched'' model for all trap species.
The lines in both panels show the predictive model (eq.~\ref{eq:sumpred}). 
We study the worst affected objects (end of the mission,
furthest from the readout register) and the faintest \textit{Euclid}
galaxies.}
\label{fig:flux2}
\end{figure}

The lower plot of Fig.~\ref{fig:flux2} shows the flux bias 
$\Delta F_{\mathrm{rel}}$ in dependence of a change $\Delta\tau_{i}$
in the correction model, like in Sect.~\ref{sec:tau}.
As for varying the $\rho_{i}$ (Sect.~\ref{sec:fluxrho}), a change
to only the fast and/or the medium traps yields only small departures from the
zeropoint such that we can bundle together all trap species for deriving a
tolerance range. The respective measurements (black circles in Fig.~\ref{fig:flux2})
show a steep slope at $\Delta\tau_{i}\!<\!0$ 
that flattens out to $\Delta\tau_{i}\!>\!0$.
This can be explained given the saturation of $\Delta F_{\mathrm{rel}}$
found at large $\tau$ in Fig.~\ref{fig:raw4panel} and is confirmed by
our model (eq.~\ref{eq:taupred}; dotted line in Fig.~\ref{fig:flux2}). 
Our prediction is offset from the measurement
due to uncertainties in the modelling, but the slopes agree well.

Although polynomial fits using eq.~(\ref{eq:polynom}) warrant cubic terms
in both cases, $\Delta F_{\mathrm{rel}}(\tau_{i}\!+\!\Delta\tau_{i})$ 
is much straighter in the 
``first pixel matched'' case where also the 
$\rho_{i}$ are altered (star symbols in Fig.~\ref{fig:flux2}; cf.\ 
Sect.~\ref{sec:sti}). The reason is that the slopes of 
$\Delta F_{\mathrm{rel}}(\rho_{i}\!+\!\Delta\rho_{i})$ and 
$\Delta F_{\mathrm{rel}}(\tau_{i}\!+\!\Delta\tau_{i})$ 
have the same sign and do not partially cancel each other out, as is the case for 
$\Delta e_{1}(\rho_{i}\!+\!\Delta\rho_{i})$ and 
$\Delta e_{1}(\tau_{i}\!+\!\Delta\tau_{i})$. 
Again, eq.~(\ref{eq:sumpred}) succeeds in predicting the measurements, despite 
offsets that are significant given the small uncertainties but small in terms
of $\Delta F_{\mathrm{rel}}$ in the uncorrected images.

Using the cubic fits, we find the following
wide tolerance ranges (eq.~\ref{eq:tol}) 
$\Delta\tau_{3,\mathrm{min}}^{\mathrm{tol}}/\tau_{3}\!=\!15.7\pm0.2$~\% and
$\Delta\tau_{3,\mathrm{max}}^{\mathrm{tol}}/\tau_{3}\!=\!23.2_{-0.9}^{+1.0}$~\%. 
In the ``first pixel matched'', case 
the intervals are considerably tighter, due to the contribution from the 
change in densities, with 
$\Delta\tau_{i,\mathrm{min}}^{\mathrm{tol}}/\tau_{i}\!=\!2.625\pm0.007$~\% and
$\Delta\tau_{i,\mathrm{max}}^{\mathrm{tol}}/\tau_{i}\!=\!2.630\pm0.007$~\%. 
Again, the strictest
constraints come from the ellipticity component $\Delta e_{1}$.

\section{Discussion} \label{sec:disc}

\subsection{Limits of the predictive model}

We measured tolerance ranges for changes in the $\rho_{i}$ and $\tau_{i}$ 
given the \textit{Euclid} VIS requirements, and presented a model 
(Eq.~\ref{eq:sumpred}) capable of predicting these results based on the 
$\Delta\eta(\tau)$ curves (e.g.\ $\Delta e_{1}(\tau)$, Fig.~\ref{fig:raw4panel}), 
that are less expensive to obtain in terms of CPU time.
However, as can be seen in particular in Fig.~\ref{fig:mdm2_exp3}, there is
a mismatch between predictions and measurements for $\tau_{3}$, 
the most common baseline model trap species. 
As discussed in Sect.~\ref{sec:sti}, this is caused by the
finite sampling and the empirical nature of eq.~(\ref{eq:adg}). 

Unfortunately, $f(\tau)$ and $f^{\mathrm{resid}}(\tau)$ 
will likely depend non-trivially on the source profile.
Moreover, Eq.~(\ref{eq:sumpred}), 
if applied to ellipticity, treats it as additive. Where this
approximation breaks down, i.e.\ when values that are not $\ll\!1$
are involved, the correct additional formula \citep[e.g.][]{2006glsw.book.....S}
must be used. This applies to
CTI-induced ellipticity as well as to large intrinsic or shear components.

We tested that the dependence on $\beta$ 
(Fig.~\ref{fig:mdm2_exp0}) can be included in the model as well, yielding
\begin{multline} \label{eq:betapred}
\Delta f^{\mathrm{Pr}}(\beta,\rho_{i},\tau_{i})\!=\!\sum_{i}\rho_{i}f^{\mathrm{resid}}(\tau_{i}) 
+ [f(\beta\!+\!\Delta\beta)\!-\!f(\beta)] \\
\times\sum_{i}\left[\rho_{i}f(\tau_{i})-(\rho_{i}\!+\!\Delta\rho_{i})f(\tau_{i}\!+\!\Delta\tau_{i})\right].
\end{multline}

\subsection{Applicability}

Our findings pertain specifically to CTI correction employing
the \citet{2014MNRAS.439..887M} iterative correction scheme, the current
nominal procedure for \textit{Euclid} VIS. Other algorithms for the removal 
of CTI trailing exist that might not be susceptible in the same way to readout noise.
\citet{2012MNRAS.419.2995P}, investigating the full-forward approach designed for
\textit{Gaia}, did not observe a readout noise floor similar to the one we found.
The same might hold for including CTI correction in a forward-modelling
shear measurement pipeline \citep[e.g.][]{2013MNRAS.429.2858M}. However, the
\textit{Gaia} method has not been applied yet to actual observational data, 
and the \citet{2014MNRAS.439..887M} is the most accurate method for the CTI
correction of real data today.

We remind the reader that our results on the zeropoints upon correcting with the
correct model (Fig.~\ref{fig:mdm2_exp0}) are dependent on the specifics of the
small and faint galaxies we simulated. Further tests will determine if the
large bias in $R^{2}$ persists under more realistic scenarios.

The narrow tolerances of $\Delta\rho/\rho\!=\!0.11$\% 
and $\Delta\tau/\tau\!=\!0.17$\% for the 
density of the slow traps species might look
daunting, but fortunately, due to the discernible trails caused by these traps
it is also the easiest species of which to determine the properties.
Conversely, the $\Delta\rho/\rho\!=\!3$\% and
$\Delta\tau/\tau\!=\!8$\% for the fast traps 
are much larger, but constraints on these traps will be harder to achieve from 
laboratory and in-flight calibration data.
Considering the ``first pixel matched'' case, taking into account how trap
parameters are determined from CTI trails, relaxes the tolerances
from ellipticity but tightens the (much broader) tolerances from the 
photometric, for our particular baseline trap mix. 
We notice that, while trap parameters are degenerate and Sect.~\ref{sec:sti}
marks a first attempt to disentangle these parameters, each (degenerate) set
of parameters can yield a viable CTI correction. Characterising the true trap
species, however, is crucial with respect to device physics applications.

Source profile-dependence of the CTI-induced flux bias $\Delta F_{\mathrm{rel}}$
will lead to a sample of realistic sources (i.e.\ with a distribution of source
profiles) showing a range in $\Delta F_{\mathrm{rel}}$ at any given readout noise
level. Thus, the uncertainty in $\Delta F_{\mathrm{rel}}$ will be larger than the $10^{-4}$
we measured for our broad-winged, but homogeneous images in Sect.~\ref{sec:fluxrn}.
More sophisticated simulations are necessary to assess the role of the variable
CTI-induced flux bias in \textit{Euclid}'s photometric error budget.

\section{Conclusions and Outlook} \label{sec:conclusion}

The goal was to bridge the divide between engineering measurements of 
CTI, and its degradation of scientific measurements 
of galaxy shapes. We have developed a very fast algorithm to model
CTI in irradiated e2v Technologies CCD273 devices, 
reproducing laboratory measurements performed at ESTEC. 
We take a worst-case approach and simulate the faintest 
galaxies to be studied by \textit{Euclid}, with a broad-winged exponential
profile, at the end of the mission and furthest away from the readout register.
Our analysis is hindered by the divergent surface brightness moments of the
Marsaglia-Tin distribution that the ellipticity components follow.
We alleviate this problem 
by means of a Taylor expansion around the mean of the denominator,
yielding an accuracy of $\sigma e_{\alpha}\!\approx\!10^{-4}$ 
by averaging over $10^{7}$ simulated images. 
We advocate that \emph{Euclid} requirements be re-defined in a way that avoids
ratios of noisy quantities. 

Our detailed study of the trapping process has confirmed 
that not all traps are equally bad for shape measurement
\citep{2010PASP..122..439R}: 
Traps with release timescales of a few 
clocking cycles cause the largest spurious ellipticity, while all traps with
longer $\tau_{i}$ yield the strongest flux bias.

The impact of uncertainties in the trap densities $\rho_{i}$ 
and time scales $\tau_{i}$ on CTI effects 
can be predicted to a satisfactory accuracy by a model that is
linear in the $\rho_{i}$ and additive in the effects of different trap species.
For future applications, this will allow
us to reduce the simulational effort
in CTI forecasts, calculating the effect of trap pixels 
from single species data. 

Informed by laboratory data of the irradiated CCD273, 
we have adopted a baseline trap model for \textit{Euclid} VIS forecasts.
We corrected images with a trap model $\mathbf{M}+\delta\mathbf{M}$ offset from 
the model $\mathbf{M}$ used for applying CTI. Thus we derived
tolerance ranges for the uncertainties in the trap parameters, given 
\textit{Euclid} requirements, positing that the required level of 
correction will be achieved. We conclude:
\begin{description}
\item[1.] In the absence of readout noise, perfect CTI correction in terms of
ellipticity and flux can be achieved.
\item[2.] Given the nominal rms readout noise of $4.5$ electrons, we measure 
$Z_{\mathrm{e_{1}}}\!=\!\Delta e_{1}\!=\!-1.18\times10^{-3}$ after CTI correction.
This still exceeds the \textit{Euclid} requirement of 
$\left|\Delta e_{1}\right|\!<\!1.1\times10^{-4}$. 
The requirement may still be met on the actual ensemble of galaxies 
\textit{Euclid} will measure, since we consider only the smallest galaxies of $S/N\!=\!11$.
Likewise, in $S/N\!=\!200$ stars, we measure a size bias of
$1.12\times10^{-3}$, exceeding the requirement of 
$\left|\Delta R^{2}/R^{2}_{\mathrm{true}}\right|\!<\!4\times10^{-4}$. 
\item[3.] The spurious ellipticity $\Delta e_{1}$ sensitively depends on the
correct well fill power $\beta$, which we need to constrain to an accuracy of 
$\Delta\beta_{\mathrm{tol}}\!=\!(6.31\pm0.07)\times 10^{-5}$ to meet requirements. 
This assumes calibration by a single, bright charge injection line. 
The narrowest tolerance intervals are found for the dominant slow trap species
in our baseline mix: 
$\Delta\rho_{\mathrm{tol}}/\rho_{0}\!=\!(\pm0.0272\pm0.0005)$\%,
and $\Delta\tau_{\mathrm{tol}}/\tau_{0}\!=\!(\pm0.0400\pm0.004)$\%. 
\item[4.] Given the nominal rms readout noise, we measure a flux bias 
$Z_{\mathrm{F}}\!=\!\Delta F_{\mathrm{rel}}\!=\!(-1.980\pm0.012)$\% 
after CTI correction, within the required
$\left|\Delta F_{\mathrm{rel}}\right|\!<\!0.7$~\% for the photometric
uncertainty. More relevant for \textit{Euclid} will be the uncertainty of this bias,
which for realistic sources depends on their source profile. Further study
is necessary here, as well as for the impact of CTI on photometric nonlinearity.
\end{description}

The final correction will only be as good as on-orbit characterisation of 
physical parameters such as trap locations, density and release time.
The next steps building on this study should include:
1.) Researching and testing novel algorithms mitigating the effects of read 
noise as part of the CTI correction.
2.) Characterising
the effect of realistic source profile distributions in
terms of the photometric and nonlinearity requirements.
3.) Translating the tolerances in trap model parameters into recommendations
of calibration measurements and their analysis, based on modelling the
characterisation of trap species. 

Plans for \textit{Euclid} VIS calibration have already been updated to 
include charge injection at multiple levels such that $\beta$ does not need
to be extrapolated from bright charge injection lines to faint galaxies. 
We will continue to liaise between engineers and scientists to determine how
accurately it will be necessary to measure these physical parameters.
The VIS readout electronics will be capable of several new
in-orbit calibration modes such as trap pumping \citep{2012SPIE.8453E..17M} 
that are not possible with HST,
and our calculations will advise what 
will be required, and how frequently they need to be 
performed, in order to adequately characterise the instrument for scientific success.

\section*{Acknowledgements}

This work used the DiRAC Data Centric system at Durham University, operated by 
the Institute for Computational Cosmology on behalf of the STFC DiRAC HPC 
Facility (www.dirac.ac.uk). This equipment was funded by BIS National 
E-infrastructure capital grant ST/K00042X/1, STFC capital
grants ST/H008519/1 and ST/K00087X/1, STFC  
DiRAC Operations grant ST/K003267/1 and Durham University. 
DiRAC is part of the National E-Infrastructure.

HI thanks Lydia Heck and Alan Lotts for friendly and helpful 
system administration.
HI acknowledges support through European Research Council
grant MIRG-CT-208994.
RM and HI are supported 
by the Science and Technology Facilities Council (grant numbers ST/H005234/1 and ST/N001494/1) 
and the Leverhulme Trust (grant number PLP-2011-003).
JR was supported by JPL, run under a contract for NASA by Caltech. 
OC and OM acknowledge support from the German Federal Ministry for Economic
Affairs and Energy (BMWi) provided via DLR under project no.~50QE1103.

The authors thank Henk Hoekstra, Peter Schneider, Yannick Mellier, 
Tom Kitching, Reiko Nakajima, Massimo Viola, and the members of 
\textit{Euclid} CCD Working group, \textit{Euclid} OU-VIS and OU-SHE groups 
for comments on the text and useful discussions.

\bibliographystyle{mn2e}
\bibliography{CTICorr_v13}

\begin{thebibliography}{37}
\expandafter\ifx\csname natexlab\endcsname\relax\def\natexlab#1{#1}\fi

\bibitem[{{Anderson} \& {Bedin}(2010)}]{2010PASP..122.1035A}
{Anderson} J., {Bedin} L.~R., 2010, \pasp, 122, 1035

\bibitem[{{Bertin} \& {Arnouts}(1996)}]{1996A&AS..117..393B}
{Bertin} E., {Arnouts} S., 1996, \aaps, 117, 393

\bibitem[{{Bristow}(2003)}]{2003astro.ph.10714B}
{Bristow} P., 2003, Instrument Science Report CE-STIS-2003-001; ArXiv
  astro-ph/0310714

\bibitem[{{Casella} \& {Berger}(2002)}]{casella+berger:2002}
{Casella} G., {Berger} R.~L., 2002, Statistical Inference, 2nd edn. Duxbury

\bibitem[{{Cropper} {et~al}\mbox{.}(2013){Cropper}, {Hoekstra}, {Kitching},
  {Massey}, {Amiaux}, {Miller}, {Mellier}, {Rhodes}, {Rowe}, {Pires}, {Saxton},
  \& {Scaramella}}]{2013MNRAS.431.3103C}
{Cropper} M. {et~al.}, 2013, \mnras, 431, 3103

\bibitem[{{Cropper} {et~al}\mbox{.}(2014){Cropper}, {Pottinger}, {Niemi},
  {Denniston}, {Cole}, {Szafraniec}, {Mellier}, {Berth{\'e}}, {Martignac},
  {Cara}, {di Giorgio}, {Sciortino}, {Paltani}, {Genolet}, {Fourmand},
  {Charra}, {Guttridge}, {Winter}, {Endicott}, {Holland}, {Gow}, {Murray},
  {Hall}, {Amiaux}, {Laureijs}, {Racca}, {Salvignol}, {Short}, {Lorenzo
  Alvarez}, {Kitching}, {Hoekstra}, \& {Massey}}]{2014SPIE.9143E..0JC}
{Cropper} M. {et~al.}, 2014, in Proc. SPIE, Vol. 9143, 9143 0J

\bibitem[{{Endicott} {et~al}\mbox{.}(2012){Endicott}, {Darby}, {Bowring},
  {Burt}, {Eaton}, {Grey}, {Swindells}, {Wheeler}, {Duvet}, {Cropper},
  {Walton}, {Holland}, {Murray}, \& {Gow}}]{2012SPIE.8453E..04E}
{Endicott} J. {et~al.}, 2012, in Proc. SPIE, Vol. 8453, 8453 03

\bibitem[{{Kacprzak} {et~al}\mbox{.}(2012){Kacprzak}, {Zuntz}, {Rowe},
  {Bridle}, {Refregier}, {Amara}, {Voigt}, \& {Hirsch}}]{2012MNRAS.427.2711K}
{Kacprzak} T., {Zuntz} J., {Rowe} B., {Bridle} S., {Refregier} A., {Amara} A.,
  {Voigt} L., {Hirsch} M., 2012, \mnras, 427, 2711

\bibitem[{{Kaiser}, {Squires} \& {Broadhurst}(1995){Kaiser}, {Squires}, \&
  {Broadhurst}}]{1995ApJ...449..460K}
{Kaiser} N., {Squires} G., {Broadhurst} T., 1995, \apj, 449, 460

\bibitem[{{Kitching} {et~al}\mbox{.}(2012){Kitching}, {Balan}, {Bridle},
  {Cantale}, {Courbin}, {Eifler}, {Gentile}, {Gill}, {Harmeling}, {Heymans},
  {Hirsch}, {Honscheid}, {Kacprzak}, {Kirkby}, {Margala}, {Massey}, {Melchior},
  {Nurbaeva}, {Patton}, {Rhodes}, {Rowe}, {Taylor}, {Tewes}, {Viola},
  {Witherick}, {Voigt}, {Young}, \& {Zuntz}}]{2012MNRAS.423.3163K}
{Kitching} T.~D. {et~al.}, 2012, \mnras, 423, 3163

\bibitem[{{Laureijs} {et~al}\mbox{.}(2011){Laureijs}, {Amiaux}, {Arduini},
  {Augu{\`e}res}, {Brinchmann}, {Cole}, {Cropper}, {Dabin}, {Duvet}, {Ealet},
  \& et~al.}]{2011arXiv1110.3193L}
{Laureijs} R. {et~al.}, 2011, ArXiv astro-ph.CO/1110.3193

\bibitem[{{Leauthaud} {et~al}\mbox{.}(2007){Leauthaud}, {Massey}, {Kneib},
  {Rhodes}, {Johnston}, {Capak}, {Heymans}, {Ellis}, {Koekemoer}, {Le
  F{\`e}vre}, {Mellier}, {R{\'e}fr{\'e}gier}, {Robin}, {Scoville}, {Tasca},
  {Taylor}, \& {Van Waerbeke}}]{2007ApJS..172..219L}
{Leauthaud} A. {et~al.}, 2007, \apjs, 172, 219

\bibitem[{{Lindegren} {et~al}\mbox{.}(2008){Lindegren}, {Babusiaux},
  {Bailer-Jones}, {Bastian}, {Brown}, {Cropper}, {H{\o}g}, {Jordi}, {Katz},
  {van Leeuwen}, {Luri}, {Mignard}, {de Bruijne}, \&
  {Prusti}}]{2008IAUS..248..217L}
{Lindegren} L. {et~al.}, 2008, in IAU Symposium, Vol. 248, IAU Symposium, {Jin}
  W.~J., {Platais} I., {Perryman} M.~A.~C., eds., pp. 217--223

\bibitem[{{Markwardt}(2009)}]{2009ASPC..411..251M}
{Markwardt} C.~B., 2009, in ASP Conference Series, Vol. 411, Astronomical Data
  Analysis Software and Systems XVIII, {D.~A.~Bohlender, D.~Durand, \&
  P.~Dowler}, ed., pp. 251--255

\bibitem[{{Marsaglia}(1965)}]{Marsaglia65}
{Marsaglia} G., 1965, Journal of the American Statistical Association, 60, 193

\bibitem[{Marsaglia(2006)}]{Marsaglia:2006:JSSOBK:v16i04}
Marsaglia G., 2006, Journal of Statistical Software, 16, 1

\bibitem[{{Massey}(2010)}]{2010MNRAS.409L.109M}
{Massey} R., 2010, \mnras, 409, L109

\bibitem[{{Massey} {et~al}\mbox{.}(2013){Massey}, {Hoekstra}, {Kitching},
  {Rhodes}, {Cropper}, {Amiaux}, {Harvey}, {Mellier}, {Meneghetti}, {Miller},
  {Paulin-Henriksson}, {Pires}, {Scaramella}, \&
  {Schrabback}}]{2013MNRAS.429..661M}
{Massey} R. {et~al.}, 2013, \mnras, 429, 661

\bibitem[{{Massey} {et~al}\mbox{.}(2014){Massey}, {Schrabback}, {Cordes},
  {Marggraf}, {Israel}, {Miller}, {Hall}, {Cropper}, {Prod'homme}, \& {Matias
  Niemi}}]{2014MNRAS.439..887M}
{Massey} R. {et~al.}, 2014, \mnras, 439, 887

\bibitem[{{Massey} {et~al}\mbox{.}(2010){Massey}, {Stoughton}, {Leauthaud},
  {Rhodes}, {Koekemoer}, {Ellis}, \& {Shaghoulian}}]{2010MNRAS.401..371M}
{Massey} R., {Stoughton} C., {Leauthaud} A., {Rhodes} J., {Koekemoer} A.,
  {Ellis} R., {Shaghoulian} E., 2010, \mnras, 401, 371

\bibitem[{{Melchior} \& {Viola}(2012)}]{2012MNRAS.424.2757M}
{Melchior} P., {Viola} M., 2012, \mnras, 424, 2757

\bibitem[{{Miller} {et~al}\mbox{.}(2013){Miller}, {Heymans}, {Kitching}, {van
  Waerbeke}, {Erben}, {Hildebrandt}, {Hoekstra}, {Mellier}, {Rowe}, {Coupon},
  {Dietrich}, {Fu}, {Harnois-D{\'e}raps}, {Hudson}, {Kilbinger}, {Kuijken},
  {Schrabback}, {Semboloni}, {Vafaei}, \& {Velander}}]{2013MNRAS.429.2858M}
{Miller} L. {et~al.}, 2013, \mnras, 429, 2858

\bibitem[{{Mor{\'e}}(1978)}]{1978LNM..630..105M}
{Mor{\'e}} J.~J., 1978, in Lecture Notes in Mathematics, Vol. 630, Numerical
  Analysis, {Springer-Verlag Berlin Heidelberg}, pp. 105--116

\bibitem[{{Murray} {et~al}\mbox{.}(2012){Murray}, {Holland}, {Gow}, {Hall},
  {Tutt}, {Burt}, \& {Endicott}}]{2012SPIE.8453E..17M}
{Murray} N.~J., {Holland} A.~D., {Gow} J.~P.~D., {Hall} D.~J., {Tutt} J.~H.,
  {Burt} D., {Endicott} J., 2012, in Proc. SPIE, Vol. 8453, 8453 16

\bibitem[{{Prod'homme} {et~al}\mbox{.}(2012){Prod'homme}, {Holl}, {Lindegren},
  \& {Brown}}]{2012MNRAS.419.2995P}
{Prod'homme} T., {Holl} B., {Lindegren} L., {Brown} A.~G.~A., 2012, \mnras,
  419, 2995

\bibitem[{{Prod'homme} {et~al}\mbox{.}(2014){Prod'homme}, {Verhoeve}, {Kohley},
  {Short}, \& {Boudin}}]{2014P1P}
{Prod'homme} T., {Verhoeve} P., {Kohley} R., {Short} A., {Boudin} N., 2014, in
  Proc. SPIE, Vol. 9154, 9154 09

\bibitem[{{Refregier} {et~al}\mbox{.}(2010){Refregier}, {Amara}, {Kitching},
  {Rassat}, {Scaramella}, {Weller}, \& {Euclid Imaging
  Consortium}}]{2010arXiv1001.0061R}
{Refregier} A., {Amara} A., {Kitching} T.~D., {Rassat} A., {Scaramella} R.,
  {Weller} J., {Euclid Imaging Consortium} f.~t., 2010, ArXiv
  astro-ph.IM/1001.0061

\bibitem[{{Refregier} {et~al}\mbox{.}(2012){Refregier}, {Kacprzak}, {Amara},
  {Bridle}, \& {Rowe}}]{2012MNRAS.425.1951R}
{Refregier} A., {Kacprzak} T., {Amara} A., {Bridle} S., {Rowe} B., 2012,
  \mnras, 425, 1951

\bibitem[{{Rhodes} {et~al}\mbox{.}(2010){Rhodes}, {Leauthaud}, {Stoughton},
  {Massey}, {Dawson}, {Kolbe}, \& {Roe}}]{2010PASP..122..439R}
{Rhodes} J., {Leauthaud} A., {Stoughton} C., {Massey} R., {Dawson} K., {Kolbe}
  W., {Roe} N., 2010, \pasp, 122, 439

\bibitem[{{Rhodes}, {Refregier} \& {Groth}(2001){Rhodes}, {Refregier}, \&
  {Groth}}]{2001ApJ...552L..85R}
{Rhodes} J., {Refregier} A., {Groth} E.~J., 2001, \apjl, 552, L85

\bibitem[{{Schneider}(2006)}]{2006glsw.book.....S}
{Schneider} P., 2006, {"Weak Gravitational Lensing" in: Gravitational Lensing:
  Strong, Weak and Micro: Saas-Fee Advanced Courses, Volume 33 }.
  Springer-Verlag Berlin Heidelberg, p. 269 ff.

\bibitem[{{Short} {et~al}\mbox{.}(2013){Short}, {Crowley}, {de Bruijne}, \&
  {Prod'homme}}]{2013MNRAS.430.3078S}
{Short} A., {Crowley} C., {de Bruijne} J.~H.~J., {Prod'homme} T., 2013, \mnras,
  430, 3078

\bibitem[{{Short} {et~al}\mbox{.}(2014){Short}, {Barry}, {Berthe}, {Boudin},
  {Boulade}, {Cole}, {Cropper}, {Duvet}, {Endicott}, {Gaspar Venancio}, {Gow},
  {Guttridge}, {Hall}, {Holland}, {Israel}, {Kohley}, {Laureijs}, {Lorenzo
  Alvarez}, {Martignac}, {Maskell}, {Massey}, {Murray}, {Niemi}, {Pool},
  {Pottinger}, {Prod'homme}, {Racca}, {Salvignol}, {Suske}, {Szafraniec},
  {Verhoeve}, {Walton}, \& {Wheeler}}]{2014SPIE.9154E..0RS}
{Short} A.~D. {et~al.}, 2014, in Proc. SPIE, Vol. 9154, 9154 0R

\bibitem[{{Smith} {et~al}\mbox{.}(2012){Smith}, {Anderson}, {Armstrong},
  {Avila}, {Bedin}, {Chiaberge}, {Davis}, {Ferguson}, {Fruchter}, {Golimowski},
  {Grogin}, {Hack}, {Lim}, {Lucas}, {Maybhate}, {McMaster}, {Ogaz}, {Suchkov},
  \& {Ubeda}}]{2012AAS...21924101S}
{Smith} L.~J. {et~al.}, 2012, in American Astronomical Society Meeting
  Abstracts, Vol. 219, American Astronomical Society Meeting Abstracts, p.
  241.01

\bibitem[{{Tin}(1965)}]{Tin65}
{Tin} M., 1965, Journal of the American Statistical Association, 60, 294

\bibitem[{{Verhoeve} {et~al}\mbox{.}(2014){Verhoeve}, {Prod'homme},
  {Oosterbroek}, {Boudin}, \& {Duvet}}]{2014V1V}
{Verhoeve} P., {Prod'homme} T., {Oosterbroek} T., {Boudin} N., {Duvet} L.,
  2014, in Proc. SPIE, Vol. 9154, 9154 16

\bibitem[{{Viola}, {Kitching} \& {Joachimi}(2014){Viola}, {Kitching}, \&
  {Joachimi}}]{2014MNRAS.439.1909V}
{Viola} M., {Kitching} T.~D., {Joachimi} B., 2014, \mnras, 439, 1909

\end{thebibliography}

\appendix

\section{Informing the baseline model with laboratory data} \label{sec:labdata}

\subsection{EPER/FPR data with irradiated CCD}

\begin{figure}
\includegraphics[width=85mm]{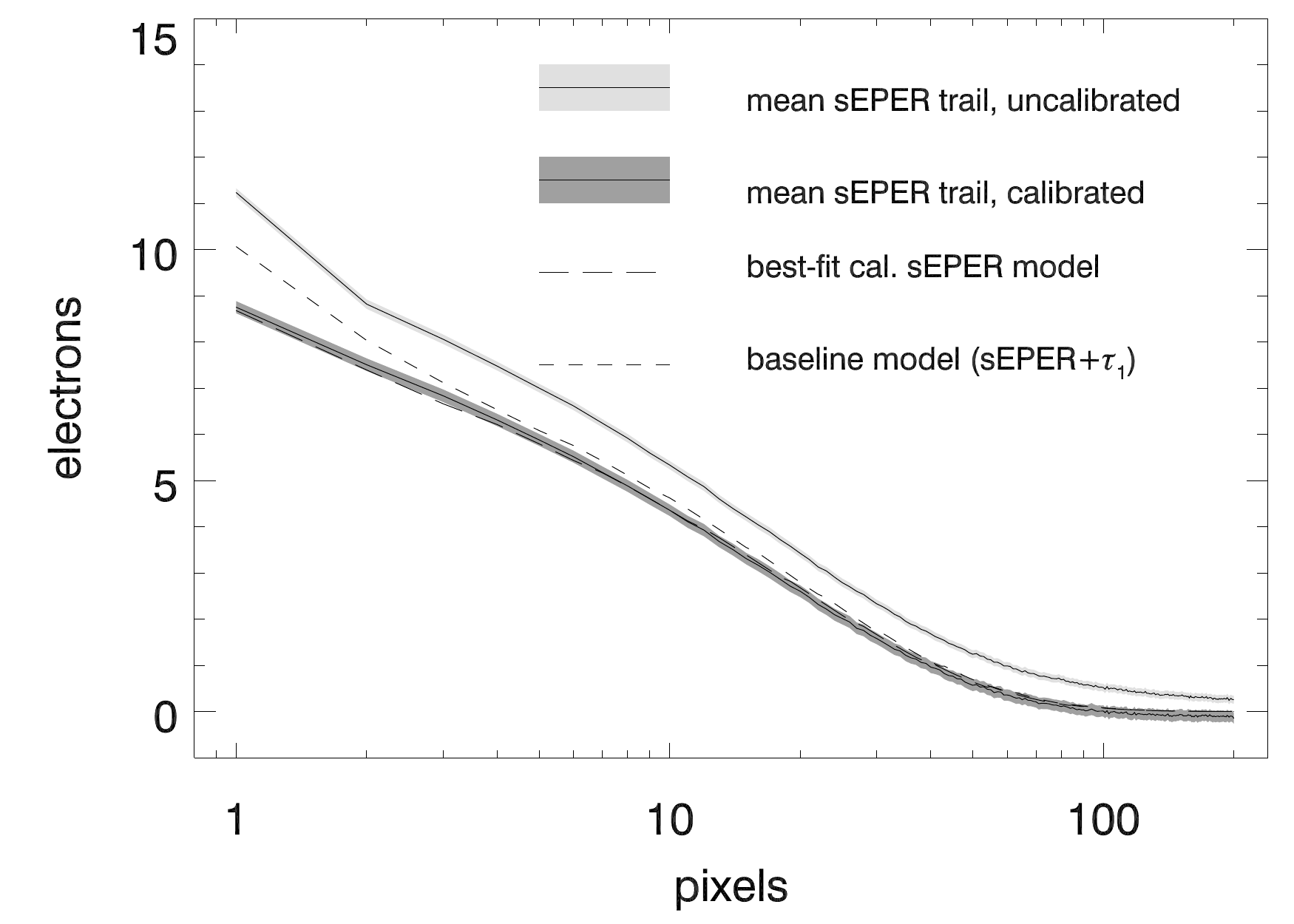}\\
\includegraphics[width=85mm]{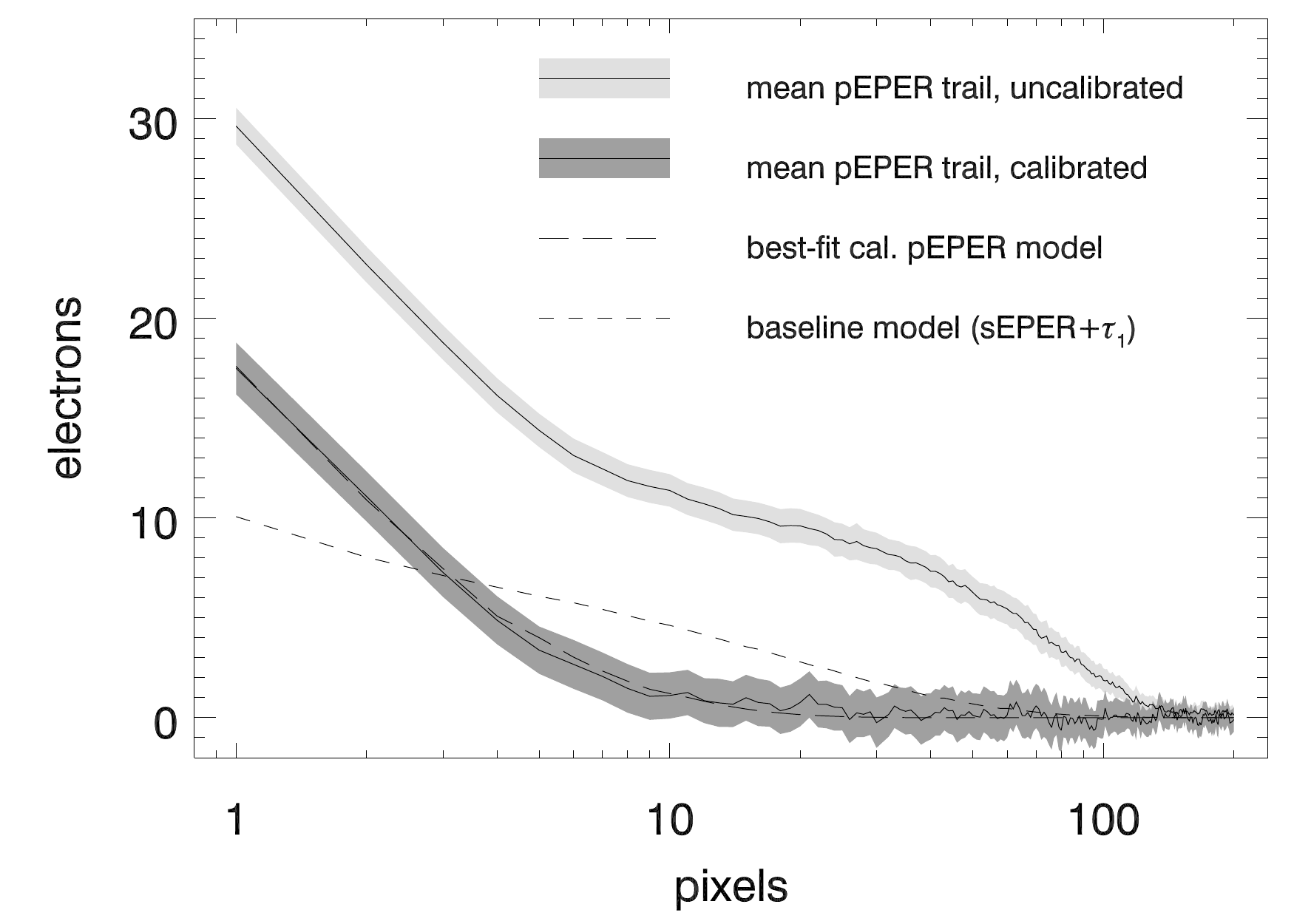}
\caption{CCD273 EPER trails in the serial (\textit{upper plot}) and parallel
(\textit{lower plot}) directions. Shown here are the G quadrant trails at
an input signal of $\sim\!1000$ electrons. Solid lines within the light and dark grey
shaded areas denote the average and its uncertainty of the profile before
and after correction for electronic effects. The best-fit model to the
corrected trail is shown as a long-dashed line. For the purpose of 
illustration, the baseline trap model is shown in both plots as a short-dashed
line. Building on the serial EPER model, the baseline model includes fast traps
that are seen in quadrant F.}
\label{fig:trails}
\end{figure}
\begin{figure}
\includegraphics[width=85mm]{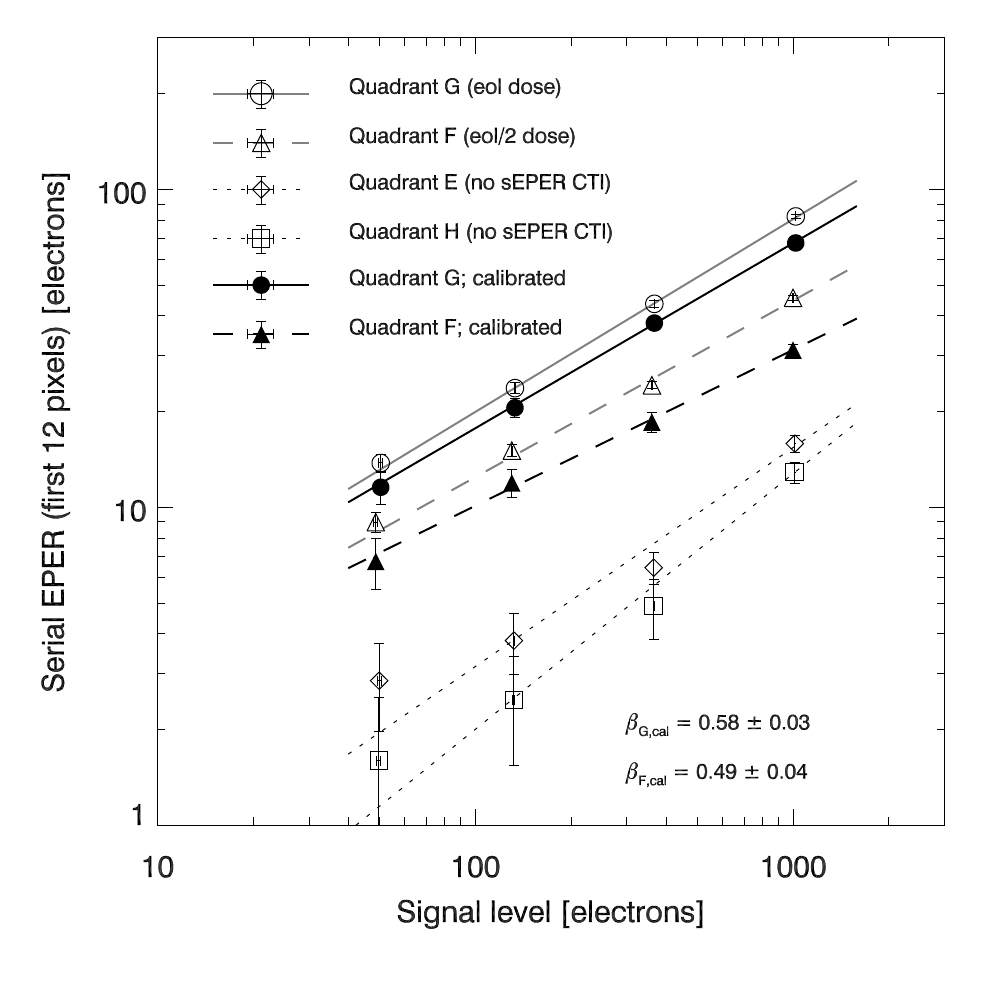}\\
\includegraphics[width=85mm]{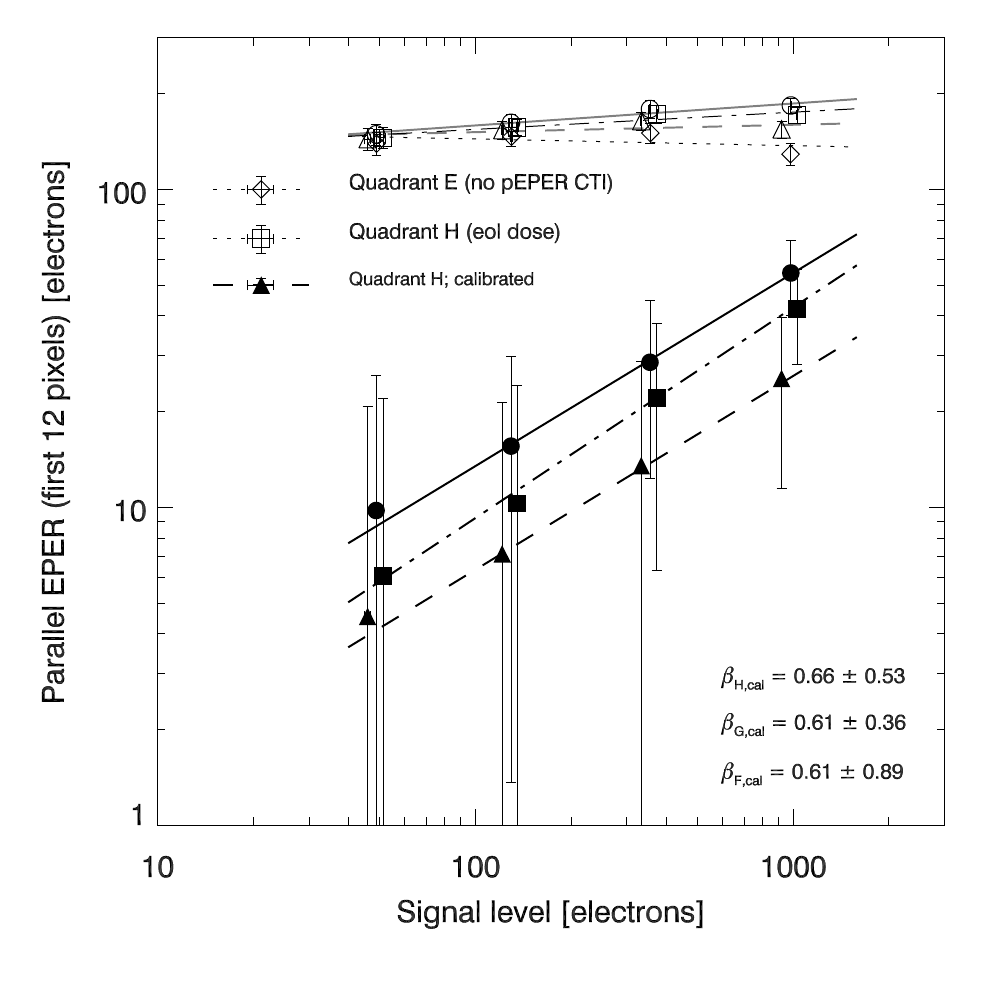}
\caption{The well fill power $\beta$ measured from the integrated EPER CTI
as a function of input signal. The \textit{upper panel} shows the results
from the serial EPER measurements, for which CTI is present in the F and G
quadrants and can be corrected using the E and H quadrants. 
The \textit{lower panel} shows the results
from the parallel EPER measurements, for which CTI is present in the F, G,
and H quadrants and can be corrected using the E quadrant. Open symbols
denote the raw measurements, filled symbols the calibrated measurements
from which the fits for $\beta$ are derived.}
\label{fig:labbeta}
\end{figure}
In this Appendix, we define a baseline CTI model for {\it Euclid} VIS. 
Our model is based upon laboratory tests of an irradiated 
e2v Technologies back-illuminated \textit{Euclid} 
prototype CCD273, analysed at ESA/ESTEC \citep{2014P1P}. The device was 
irradiated at ambient room temperature using $10.4$~MeV protons, degraded from 
a $38.5$~MeV primary proton beam at the 
Kernfysisch Versneller Instituut, Groningen, in April 2013. 
Two different shielding masks were used
\citep{2014P1P} resulting in the four quadrants of the CCD, 
called E, F, G, and H, and
corresponding to the four output nodes, receiving different radiation doses.
Each a half of two quadrants, called G and H, received a 
$10$~MeV equivalent fluence of $4.8\times 10^{9}\,\mbox{protons}/\mbox{cm}^{-2}$, 
representative of the predicted end-of-life (eol) proton fluence for \textit{Euclid}.
Half of the F quadrant was irradiated with a $10$~MeV equivalent fluence of
$2.4\times 10^{9}\,\mbox{protons}/\mbox{cm}^{-2}$, the $\mbox{eol}/2$ fluence. 
Neither the E quadrant, the serial register of the H quadrant, nor the readout
nodes were irradiated \citep{2014V1V,2014P1P}. 

At the ESA Payload Technology Validation section CCD test bench located at ESTEC
\citep{2014V1V}, the irradiated CCD273 was characterised at the \textit{Euclid} VIS 
nominal conditions of $153\,\mbox{K}$ temperature and a $70\,\mbox{kHz}$
readout frequency. While a serial clocking scheme with the same width for each 
register phase at each step was used, minimising serial CTI, the nominal 
line/parallel transfer duration of $0.11\,\mbox{ms}$ was not optimised.

As part of the characterisation, a suite of extended pixel edge response (EPER)
and first pixel response (FPR) experiments were
performed, at different 
flatfield signal levels. For the purpose of deriving a fiducial baseline model 
of the charge traps present in the CCD273, we focus on the parallel and serial 
EPER data. To study the serial EPER (sEPER) CTI, a flatfield image is taken,
then the half opposite to the readout direction is dumped; then the frame is
read out. This yields a flatfield with a sharp trailing edge in flatfield signal.
Electrons captured from flatfield pixels are being released into signal-less
pixels, resulting in a CTI trail. Our parallel EPER (pEPER) tests make use of the
parallel overscan region, providing a similar signal edge.
 
Each measurement was performed repeatedly, in order to gather statistics: 
$45$ times for the sEPER data at low signal, and $20$ times for the pEPER data.
Raw trail profiles are extracted from the first $200$ pixels following the 
signal edge, taking the arithmetic mean over the independent lines 
perpendicular to the direction (serial or parallel) of interest.
The same is done in the overscan region, unexposed to light, and the
pixel-by-pixel median of this reference is subtracted as 
background from the raw trails. In the same way as the reference, the median flatfield signal
is determined, and also corrected for the overscan reference. Finally,
the trail (flatfield signal) at zero flatfield exposure time is subtracted
from the trails (flatfield signals) at exposure times $>\!0$.

Figure~\ref{fig:trails} shows the resulting ``uncalibrated'' trail profiles for 
the sEPER (upper panel) and pEPER (lower panel) measurements in the G quadrant
(eol radiation dose), at a flatfield exposure time corresponding to an average
of $1018$ signal electrons per pixel. These are the upper solid lines with
light grey shading denoting the propagated standard errors from the repeated
experiments. Effects in the readout electronics mimic CTI. 
We correct for the electronic effect by subtracting the average trail in the
unirradiated quadrants (E for pEPER, and E and H for sEPER). 
The resulting ``calibrated'' trail profiles and their uncertainties are 
presented as the lower solid lines and dark grey shadings in Fig.~\ref{fig:trails}.
The calibration makes a small correction to the sEPER trail which is dominated
by slow traps, yielding a significant signal out to $\sim\!60$ pixels.
On the contrary, the electronic effect accounts for $1/3$ of the uncalibrated
pEPER trail even in the first pixel, and for all of it beyond the tenth.
Thus the $S/N$ in the calibrated trail is much lower.

\subsection{The well fill power $\beta$}
\begin{table}
\caption{The same as Table~\ref{tab:traps}, but for the best-fit models shown 
in Fig.~\ref{fig:trails}. 
The baseline well fill power is $\beta_{0}\!=\!0.58$.}
\begin{tabular}{cccc}\hline\hline
best-fit sEPER model & $i\!=\!1$ & $i\!=\!2$ & $i\!=\!3$ \\
Trap density $\rho_{i}$ [px$^{-1}$] & $0.01$ & $0.03$ & $0.90$ \\
Release timescale $\tau_{i}$ [px] & $0.8$ & $3.5$ & $20.0$ \\ \hline
best-fit pEPER model & $i\!=\!1$ & $i\!=\!2$ & $i\!=\!3$ \\
Trap density $\rho_{i}$ [px$^{-1}$] & $0.13$ & $0.25$ & $--$ \\
Release timescale $\tau_{i}$ [px] & $1.25$ & $4.4$ & $--$ \\
\hline\hline \label{tab:models}
\end{tabular}
\end{table}
In a volume-driven CTI model, the cloud of photoelectrons in any given
pixel is assumed to fill up a height within the silicon that increases as 
electrons are added (Eq.~\ref{eq:nenc}). 
The growth of the cloud volume is governed by the term 
$\left(\frac{n_{\mathrm{e}}}{w}\right)^{\beta}\sum_{i}\rho_{i}$ 
in Eq.~(\ref{eq:nenc}), with the full-well depth $w\!=\!84700$
limiting the maximum number of electrons in a pixel.
There is no supplementary buried channel in the CCD273, which for 
\textit{HST}/ACS leads to
the first $\sim100$ electrons effectively occupying zero volume 
\citep{2010MNRAS.401..371M}.

\begin{figure*}
\vspace{-0.2cm}
\includegraphics[width=150mm,angle=180]{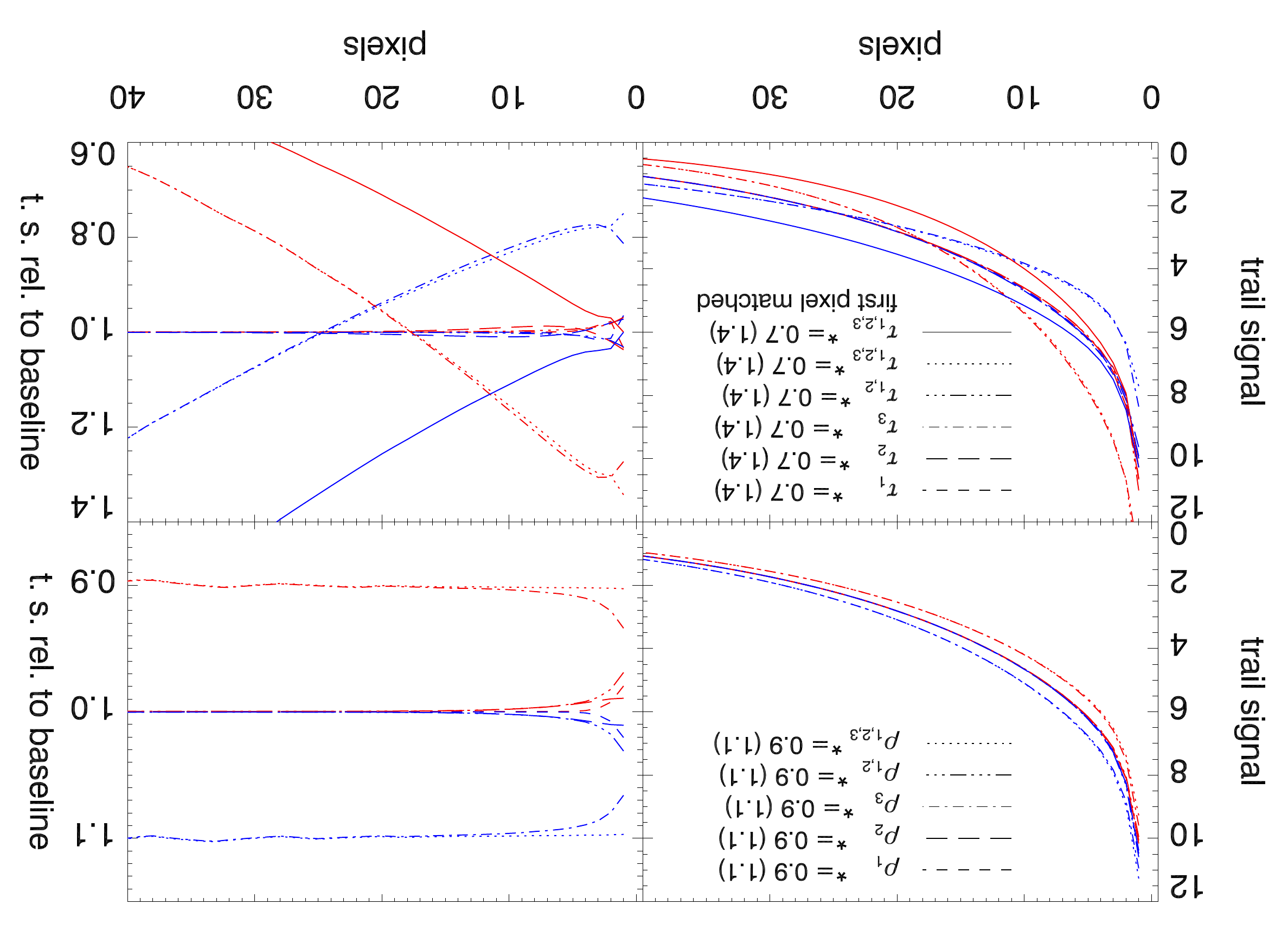}
\caption{Trail profiles corresponding to our trap models. 
\textit{Upper left panel:} Trails for models in which the
densities $\rho_{i}$ are modified by a factor 
$0.9$ (red) or $1.1$ (blue).
This is applied to: the fast traps $\rho_{1}$ (dashed lines); the medium traps 
$\rho_{2}$ (long-dashed lines); the slow traps $\rho_{3}$ (dot-dashed
lines); $\rho_{1}$ and $\rho_{2}$ (triple dot-dashed lines); all traps 
(dotted lines). \textit{Upper right panel:} The same, but relative to 
the baseline model. \textit{Lower left panel:} Trails for trap models in which the
release time scales $\tau_{i}$ are modified by a factor $0.7$ (red) 
or $1.4$ (blue). The line code is the same as above.
Solid lines show models in which not only all the $\tau_{i}$ are changed,
but also the $\rho_{i}$ adjusted such that the first pixel matches
the baseline trail profile (Sect.~\ref{sec:sti}). 
\textit{Lower right panel:} The same, but relative to the baseline model.}
\label{fig:models}
\end{figure*}
We use measurements of the integrated EPER as a function of input signal to
constrain the well fill power $\beta$ of the trapping model.
Our simulated galaxies are faint; so we restrict ourselves to the four lowest
signal levels measured in the laboratory, with up to $\sim\!1000$ electrons.
The input signal is measured as the average count difference between the 
flatfield and overscan regions, corrected for the CCD gain.

Figure~\ref{fig:labbeta} shows the CTI trails from Fig.~\ref{fig:trails},
integrated over the first $12$ pixels. We checked that integrating over 
up to the full overscan region length of $200$ pixels does not change the
results drastically.
In the sEPER data (upper panel of Fig.~\ref{fig:labbeta}), the unirradiated
quadrants E and H (open squares and diamonds) exhibit very small trail 
integrals (caused by the readout electronics); one order of magnitude smaller
than in the irradiated quadrants F and G (open circle and triangle).
Hence, calibrating out the instrumental effect by subtracting the arithmetic
average from the E and H quadrants yields only a small correction to the
F and G trail integrals. To these calibrated sEPER measurements (filled circle
and triangle), we fit linear relations in log-log-space using the 
\texttt{IDL fitexy} routine and measure $\beta_{\mathrm{F,cal}}\!=\!0.49\pm0.04$ 
and $\beta_{\mathrm{G,cal}}\!=\!0.58\pm0.03$.

We repeat this procedure for the pEPER measurements where the unirradiated
E quadrant shows a similar EPER integral than the irradiated F, G, and H
quadrants (lower panel of Fig.~\ref{fig:labbeta}). Thus, the pEPER and sEPER
integrals may yield similar values as a function of signal, but for pEPER
the low $S/N\!\!\ll\!1$ causes large uncertainties. Consequently, $\beta$ is
not well constrained, with $\beta_{\mathrm{F,cal}}\!=\!0.66\pm0.53$, 
$\beta_{\mathrm{G,cal}}\!=\!0.61\pm0.36$, and 
$\beta_{\mathrm{H,cal}}\!=\!0.61\pm0.89$, but they agree with the sEPER results.

In conclusion, we adopt a baseline well-fill power of $\beta_{0}\!=\!0.58$ for
our further tests, based on the precise sEPER result for the full radiation dose.

\subsection{From trail profiles to trap parameters} \label{sec:baseline}

To constrain the trap release time-scales $\tau_{i}$ and trap densities
$\rho_{i}$, we make use of the two signal levels of $\sim\!360$ electrons and
$\sim\!1000$ electrons that bracket the number of electrons we expect to be
found in a typical faint \textit{Euclid} galaxy. These are the two highest
data points in Fig.~\ref{fig:labbeta}.
We compare the average, measured, calibrated trails from the irradiated 
quadrants (examples for the G quadrant are presented in Fig.~\ref{fig:trails})
and compare them to the output a one-dimensional version of our
\citet{2014MNRAS.439..887M} clocking routine produces given trap densities
$\rho_{i}$ and release timescales $\tau_{i}$, and under circumstances close
to the laboratory data (i.e.\ a $200$ pixel overscan region following a
$2048$ pixel flatfield column of $1018$ signal electrons).
The fitting is performed using the \texttt{MPFIT} implementation of the
Levenberg-Marquardt algorithm for nonlinear regression
\citep{2009ASPC..411..251M,1978LNM..630..105M}.

Fitting a sum of exponentials is remarkably sensitive to noise in the data
because the parameters ($\tau_{i}$ and $\rho_{i}$) we are probing are
intrinsically correlated. We assess the robustness of our results by repeating
the fit not only for the two (three) irradiated sEPER (pEPER) quadrants at
two signal levels, but for a wide range of trail lengths ($60\!<\!K\!<\!150)$
we consider, and with and without adding a constant term.

There are several possible trap species as defined by their $\tau_{i}$ that
show up in our data set. We rule out those of very low densities and consider
the frequent ``species'' whose time-scales are within a few percent of
each other as one. 
Still, this leaves us with more than one contesting family of trap species
that yet give similar trails in some of the quadrant/signal combinations.
Because, at this stage, our goal is to derive \emph{a plausible baseline model}
rather than pinpointing the correct trap species, we filter for the most common
$\tau_{i}$ and give precedence to the higher-$S/N$ data (sEPER, end-of-life
dose, $1000$ signal electrons). 
The resulting best-fit models are shown in 
Table~\ref{tab:models} and Fig.~\ref{fig:trails}.
The actual baseline model (Table~\ref{tab:traps}; short-dashed line in 
Fig.~\ref{fig:trails}) includes additional fast traps seen in the lower-$S/N$ data. 
We raise the density from $0.94$ traps per pixels to a mnemonic total of 
$1$~trap per pixel at end-of-life dose.
More refined methods will be used to determine the trap species
in a more detailed analysis of irradiated CCD273 data.

Because only $464$ pixels of the serial register in the test device 
were irradiated, the effective density of charge traps an electron clocked 
through it experiences is smaller by a factor of $464/2051$ than the actual
trap density corresponding to the end-of-life radiation dose that was applied. 
We correct for this by quoting results that have been scaled up by a factor of
$2051/(464\times0.94)\!\approx\!4.155$. 

\subsection{Example CTI trails} \label{sec:trailexamples}

Figure~\ref{fig:models} shows, for the largest deviations from the baseline
trap model we consider, their effect on the shape of the CTI trails.
Using our CTI model, we simulated the trail caused by a single pixel containing
a signal of $\sim\!1000$ electrons, comparable to a hot pixel in actual CCD data.

\label{lastpage}

\end{document}